\documentclass[%
 reprint,
 amsmath,amssymb,
 aps,
]{revtex4-1}

\usepackage{graphicx}
\usepackage{dcolumn}
\usepackage{bm}

\begin{document}

\title{Towards a General Theory of Extremes for Observables of Chaotic Dynamical Systems}

\author{Valerio Lucarini}
\email{Email: \texttt{valerio.lucarini@uni-hamburg.de}}
\altaffiliation{Also at: Department of Mathematics and Statistics, University of Reading, Reading, UK.}\affiliation{Institute of Meteorology, Klimacampus, University of Hamburg, Grindelberg 5, 20144, Hamburg, Germany}%
\author{Davide Faranda}
\altaffiliation{Now at: Service de Physique de l'Etat Condens\'e, DSM, CEA Saclay, CNRS URA 2464,
Gif-sur-Yvette, France.}\affiliation{Institute of Meteorology, Klimacampus, University of Hamburg, Grindelberg 5, 20144, Hamburg, Germany}
\author{Jeroen Wouters}%
\affiliation{Institute of Meteorology, Klimacampus, University of Hamburg, Grindelberg 5, 20144, Hamburg, Germany}%
\author{Tobias Kuna}%
\affiliation{Department of Mathematics and Statistics, University of Reading, Reading, UK.}

%


\date{\today}

\begin{abstract}
In this paper we provide a connection between the geometrical properties of
a chaotic dynamical system and the distribution of extreme values. We show
that the extremes of so-called physical observables are distributed
according to the classical generalised Pareto distribution and
derive explicit expressions for the scaling and the shape parameter. In
particular, we derive that the shape parameter does not depend on the
chosen observables, but only on the partial dimensions of the invariant
measure on the stable, unstable, and neutral manifolds. The shape
parameter is negative and is close to zero when high-dimensional systems
are considered. This result agrees with what was derived recently using
the generalized extreme value approach. Combining the results obtained
using such physical observables and the properties of the extremes of
distance observables, it is possible to derive estimates of the partial
dimensions of the attractor along the stable and the unstable directions
of the flow. Moreover, by writing the shape parameter in terms of
moments of the extremes of the considered observable and by using linear
response theory, we relate the sensitivity to perturbations of the shape
parameter to the sensitivity of the moments, of the partial
dimensions, and of the Kaplan-Yorke dimension of the attractor.  Preliminary numerical investigations provide encouraging results on the applicability of the theory presented here. The results presented here do not apply for all combinations of Axiom A systems and observables, but the breakdown seems to be related to very special geometrical configurations.

\end{abstract}

\pacs{Valid PACS appear here}
\maketitle


\section{Introduction}

Extreme value theory (EVT) is gaining more and more prominence in a vast range of scientific fields because of its theoretical relevance in mathematical and physical sciences, 
and because it addresses the problem of understanding, modeling, evaluating risk factors such as those related to instabilities in the financial markets and to natural hazards related to seismic, climatic and hydrological extreme events. Even if the probability of extreme events is very low and decreases quickly with their magnitude, the associated risks can dominate over those coming from events belonging to the bulk of the statistics.   An extensive account of recent results and relevant applications is given in \citep{ghil2010extreme}.

The first comprehensive discussion of EVT dates back to the fundamental work by \citet{gnedenko}, who investigated the distribution of the maxima of a sample of independent identically distributed (i.i.d) stochastic variables. He showed that under very general conditions such maxima are distributed according to the so-called Generalised Extreme Value (GEV) distribution. The classic way of dealing with the statistical inference of extremes actually follows quite precisely the steps of the Gnedenko's theorem. One partitions the experimental time series into bins of fixed length, extracts the maximum of each bin, and fits these data to the GEV distribution family using, typically, methods such as maximum likelihood estimation (MLE) or L-moments. See \citep{felici1} for a detailed account of this methodology. It is possible to deal with extremes by taking a different point of view, i.e., by defining extremes as the events exceeding a given threshold. In the limit of very high threshold,  we expect that the extremes are distributed according to the Generalized Pareto Distribution (GPD) introduced by \citet{pickands1975statistical} and \citet{balkema1974residual}. In the case of i.i.d. variables, it is well known that a strong connections exists between the two methodologies. As shown in \citep{leadbetter}, we have that if block maxima obey the GEV distribution, then exceedances over some high threshold obey an associated GPD. Nonetheless, it is apparent that, whereas the two approaches are equivalent in the asymptotic limit, the GPD approach provides more reliable and more robust results when  realistic, finite time series are considered (see, e.g., \citep{ding2008newly}).

\subsection{A Brief Recapitulation of Extreme Value Theory}
\citet{gnedenko}  studied the convergence of maxima of i.i.d.
variables $$X_0, X_1, .. X_{m-1}$$ with cumulative distribution function (cdf) $F_m(x)=P\{a_m(M_m-b_m) \leq x\}$ where $a_m$ and $b_m$ are normalizing sequences and $M_m=\max\{ X_0,X_1, ..., X_{m-1}\}$. Under general hypothesis on the nature of the parent distribution of data, \citet{gnedenko} showed that the asymptotic distribution of maxima, up to an affine change of variable, belongs to a single family of generalized distribution called GEV distribution whose cdf can be written as:
\begin{equation}
\lim_{m\rightarrow\infty}F_m(x)=F_{GEV}(x;\mu,\alpha,\kappa)=\textrm{e}^{-t(x)}
\label{GEV}
\end{equation}
where
\begin{equation}
t(x) = \begin{cases}\big(1+\kappa(\tfrac{x-\mu}{\alpha})\big)^{-1/\kappa} & \textrm{if}\ \kappa\neq0 \\ e^{-(x-\mu)/\alpha} & \textrm{if}\ \kappa=0\end{cases}.
\label{GEV1}
\end{equation}
This expression holds for $1+{\kappa}(x-\mu)/\alpha>0 $, using $\mu \in \mathbb{R}$ (location parameter) and $\alpha>0$ (scale parameter) as scaling constants, and ${\kappa} \in \mathbb{R}$ is the shape parameter (also called the tail index).  
When ${\kappa} \to 0$, the distribution corresponds to a Gumbel type (Type 1 distribution).  When the index is positive, it corresponds to a Fr\'echet (Type 2 distribution); when the index is negative, it corresponds to a Weibull (Type 3 distribution).

We briefly mention the Pareto approach to EVT. We define an exceedance as $z=X-T$, which measures by how much the variable $X$ exceeds a given threshold $T$. As discussed above, under the same conditions under which the block maxima of the i.i.d. stochastic variables $X$ obey the GEV statistics, the exceedances $z$ are asymptotically distributed according to the Generalised Pareto Distribution.  Defining $F_T(z)=P(X-T<z|X>T)$, we have that $\lim_{T\rightarrow \infty} F_T(z)=F_{GPD}{(z;\xi,\sigma)}$, with \citep{pickands1975statistical,leadbetter}:
\begin{equation}
F_{GPD}{(z;\xi,\sigma)} = \begin{cases}
1- \left(1+ \frac{\xi z}{\sigma}\right)^{-1/\xi} & \text{for }\xi \neq 0, \\
1-\exp \left(-\frac{z}{\sigma}\right) & \text{for }\xi = 0,
\end{cases}\label{GPD}
\end{equation}
where the range of $z$ is $0 \leq z < \infty$ if $\xi \geq 0$ and $ 0 \leq z \leq- \sigma/\xi$ if $\xi<0$. 

The relation between GEV and GPD parameters has been already discussed in literature in case of i.i.d variables  \citep{katz2005statistics,coles2001introduction,ding2008newly,malevergne2006power}. It is first interesting to note that
\begin{equation}
 F_{GPD}(z-T; \sigma,\xi) = 1+\log\left(F_{GEV}(z; T, \sigma, \xi) \right)
 \label{con}
 \end{equation}
for $F_{GEV}(z; \mu, \alpha, \kappa)\geq \exp^{-1}$, where the latter condition implies $z\geq T$ \citep{reiss07}.  If we consider the upper range $z \gg T$, we have that  $F_{GEV}(z; T, \sigma, \xi)$ is only slightly smaller than 1, so that Eq. \ref{con} implies that 
\begin{align}
F_{GPD}(z-T; \sigma,\xi) & \sim  F_{GEV}(z; T, \sigma, \xi)\nonumber \\&+O(\left(1+ \xi (z-T)/\sigma\right)^{-2/\xi})\nonumber,
\end{align}
so that the two distributions are asymptotically equivalent. This simple result is actually equivalent to the rather cumbersome formulas given in \citet{coles2001introduction} and \citet{katz2005statistics} (and reported also by us in \citep{lucarinietal2012} for defining the correspondence between the parameters of the GPD and GEV distributions describing the statistics of extreme events extracted from the same data series. 


\subsection{Extreme Value Theory for Dynamical Systems}
Recently, a great deal of attention has focused on understanding to what extent EVT can be applied to study the extreme of observables of deterministic dynamical systems. The main applications-driven motivation for this renewed interest probably comes from the spectacular development of numerical modeling in a geophysical fluid dynamical context and from the need to assess the ability of climate model to reproduce the observed statistical properties of extremes in present climate conditions and understand how they will change in an altered climate \citep{IPCC2007}. Other related applications include the numerical simulation of hydrological risk and of the production of electric energy from wind. It is clear that the matter is far from being trivial: numerical experiments on climate models of various degrees of complexity have shown that the speed of convergence (if any) of the statistical properties of the extremes definitely depends strongly on the chosen climatic variable of interest \citep{kharin2005,vannitsem,felici1,vitolo}.

Apart from these specific albeit very relevant applications, this problem has been addressed by the mathematical and statistical physical community. A first important result is that when a dynamical system has a regular (periodic of quasi-periodic) behaviour, we do not expect, in general, to find convergence to GEV distributions for the extremes of any observable \citet{nicolis,haiman}. Instead, if one chooses specific observables  and considers dynamical systems obeying suitable mixing conditions, which guarantee the independence of subsequent maxima, it is possible to prove that the distribution of the block maxima of the observables converge to a member of the GEV family. The observables are expressed as $g(dist(x,x_0))$, a function $g$ of the distance of the orbit $x$  from a point in the attractor $x_0$, usually taken as the initial condition, such that $g(y)$ has a global maximum for $y=0$. The specific member of the GEV family (which is determined by the sign of the shape parameter) the maxima distribution converges to depends on the specific choice of $g$. The paper by  \citet{collet2001statistics} can be considered the cornerstone for the subsequent results obtained in the last few years \citep{freitas2008,freitas,gupta2009extreme}. The resulting parameters of the GEV distributions can be expressed as simple functions of the local dimension of the attractor. These results have been shown to be accurately detectable in numerical experiments when considering finite time series \citep{faranda2011generalized,faranda2011numerical,faranda2011extreme}. 
If, instead, the maxima are clustered, so that they feature a relative strong short-time correlation, the results have to be modified by introducing the \textit{extremal index}  \citep{freitas2010extremal}.

Recently, it has been shown how to obtain results which are independent on whether the underlying dynamics of the system is mixing or, instead, regular . The key ingredient relies on using the Pareto rather than the Gnedenko approach. Such a shift in the point of view on extremes allows to derive results that do not dependent on whether extremes feature strong time-correlations or not. Assuming only that the local measures scales with the local dimension \citep{Bandt2007}, it is possible to obtain by direct integration a GPD for the threshold exceedances of the observables $g(dist(x,x_0))$ introduced in \citep{freitas2008,freitas,freitas2010extremal,gupta2009extreme} when considering a generic orbit of a dynamical systems. In fact, the Pareto approach entails sampling all the points of the orbit that are very close to $x_0$, thus sampling the local scaling of the invariant measure. With a suitable choice of $g(dist(x,x_0))$, the resulting $\xi$ of the GPD is proportional to the inverse of local dimension \cite{lucarinietal2012}. The results obtained using the Pareto approach agree exactly with what was derived using the Gnedenko approach under the assumption of mixing dynamics in \citep{freitas2008,freitas,gupta2009extreme,faranda2011generalized,faranda2011numerical,faranda2011extreme}.  When the underlying system is mixing enough, the dynamical (Gnedenko approach) and the geometrical (Pareto approach) points of view on extremes give the same results, whereas differences emerge if the dynamics is such that strong time correlations exist between block-maxima. The selection of the extremes of the $g$ observables discussed above - using either the Gnedenko or the Pareto approaches -  acts as magnifying lens near the initial condition, and that's why one can extract information on the local dimension. Therefore, this provides a potentially viable alternative to e.g. the Grassberger-Procaccia algorithm \citep{grass83} for the investigation of the scaling properties of the invariant measure of a chaotic attractor. 

The results discussed above feature a major drawback when considering their relevance in many applications. Extreme events correspond to close returns of the orbit to to its initial condition. While relevant problems in natural sciences can be set in the framework of this class of observables (e.g. the classic problem of \textit{weather analogues}in meteorology, already discussed by Lorenz in connection to the problem of predictability \cite{lorenz1969}), this is not the typical structure of the observables encountered in many applications, such as the case of total energy or enstrophy of a fluid flow. Recently, this problem has been addressed in \citet{holland2012}, who have studied, using the GEV approach, whether EVT applies for \textit{physical} observables of maps obeying the mixing conditions proposed in \citet{freitas2008,freitas,freitas2010extremal,gupta2009extreme}. They consider a general observable $A=A(x)$ reaching its maximum value $A_{max}$ restricted to the support of the invariant measure in $x=x_{0}$ (assuming for simplicity that such a point is unique), and  assume that $\nabla A|_{x=x_{0}} \neq 0$. Note that $A(x)$ is indeed not of the form $g(\text{dist}(x,x_0))$ discussed above. They find that the block maxima of $A(x)$ are asymptotically distributed according to a member of the GEV distributions, where the shape parameter is negative and can be written as a simple function of the partial dimensions along the stable and unstable manifolds at the point. This seems indeed a very relevant result, as it provides a universal property of extremes, regardless of the specific functional form of the observable $A$.  

\subsection{Goals of the Paper}

In this paper, we consider a GPD approach to EVT and try to complement and improve the results presented in \citep{holland2012} regarding the physical observables and those presented in \citep{lucarinietal2012} regarding the distance observables.  We focus our attention on Axiom A systems \cite{eckmann85}, which are a special class of dynamical systems possessing a Sinai-Ruelle-Bowen (SRB) invariant measure \cite{young_what_2002} and featuring hyperbolicity in the attracting set. Such invariant measure coincides with the Kolmogorov's physical measure, \textit{i.e.} it is robust against infinitesimal stochastic perturbations. Another important property of Axiom A systems is that it is possible to develop a response theory for computing the change in the statistical properties of any observable due to small perturbations to the flow \cite{ruelle98,ruelle2009}. Such response theory has recently been the subject of intense theoretical \cite{lucarini2008,lucarini2012}, algorithmic \cite{majda07} and numerical investigations \cite{reick02,cessac_linear_2007,lucarini2009,lucarini2011} and is gaining prominence especially for geophysical fluid dynamical applications. Moreover, the response theory seems to provide powerful tools for studying multiscale systems and deriving parametrizations of the impact of the fast variables on dynamics of the slow  \citep{wouters_disentangling_2012,woutersbis2012}. Finally, an important property of Axiom A systems is that, while the dynamics of natural or artificial systems is definitely not Axiom A in general, Axiom A systems can be considered as good 'effective' models of actual systems with many degrees of freedom thanks to the so-called chaotic hypothesis, which is somewhat the equivalent in the non-equilibrium framework of the classic ergodic hypothesis for equilibrium dynamics \cite{galla95}. Moreover, as discussed in \cite{lucarini2011}, when we perform numerical simulations we implicitly assume that the system under investigation is Axiom A or Axiom A-equivalent. Therefore, considering Axiom A systems seems a good mathematical framework in view of providing results useful for a large spectrum of applications. The choice of considering Axiom A systems is instrumental in the derivation of various results on the relationship of EVT parameters to the dynamical and geometrical properties of the system, and will allow addressing the problem of the sensitivity of extremes to small perturbations of the system. The dependence of the properties of extremes of parametric modulations of the underlying dynamics  is an issue of relevant theoretical as well as applicative interest. The practical interest stems from the fact that it is relevant to be able to control or predict variations in extreme events due to small perturbations to the dynamics. The theoretical interest comes from the fact that when considering extremes, universal parametric probability distributions can be defined, as opposed to the case of the bulk statistical properties. Because of this, we may hope to reconstruct the parameters descriptive of the EVT from simple moments of the distributions, express these in terms of observables of the system, and use the Ruelle response theory for expressing rigorously the sensitivity of extremes to small perturbations to the dynamics.

In Sec. \ref{EVTAxiomA} we show that by direct integration it is possible to derive the value of the two GPD parameters $\xi$ and $\sigma$, and, in particular, that the value of $\xi$ agrees with the GEV shape parameter obtained in \cite{holland2012}. We also show that, combining the results obtained using such physical observable $A$ and the distance observables considered in \cite{lucarinietal2012}, it is possible to derive the estimates of the partial dimensions of the attractor along the stable and the unstable directions of the flow. In Sec. \ref{ResponseEVT}, we  develop a linear response theory describing the impact of small time-independent $\epsilon$-perturbations to the flow on the statistical properties of the extremes of the observable $A$. We will first investigate the sensitivity of suitable defined observables describing above-threshold $A(x)$ occurrences. We will focus on computing the changes of the shape parameter $\xi$. We will find two equivalent expression for the sensitivity of $\xi$ with respect to $\epsilon$. First, we will provide an expression for the sensitivity of $\xi$ in terms of the first two moments of the probability distribution of above-threshold $A(x)$ events. Such expression entails a combination of observables of the Axiom A system, so that one can use Ruelle's theory to compute the response to $\epsilon$-perturbations to the dynamics. Nonetheless, we show that mathematical problems emerge when considering some limits. We follow the same approach for defining an expression for the sensitivity of the extremes of the distance observables considered in \citep{lucarinietal2012}. Then, we will relate the sensitivity of $\xi$ to the sensitivity of the Kaplan-Yorke dimension of the attractor. We will link our results to the well-established fact that both the extremes of observables and quantities like the Lyapunov exponents feature a good degree of regularity with respect to perturbations when one considers intermediate complexity to high-dimensional chaotic dynamical systems. In Sec. \ref{nume} we present the results of some numerical experiments performed using simple H\'enon maps \cite{henon}, aimed at providing support to our results. In Sec. \ref{scales} we briefly discuss the problems one faces when multiple time scales are present in the system. In Sec. \ref{conclu} we comment our findings and present our conclusions.

\section{Extreme Value Theory for Physical Observables of Axiom A systems}
\label{EVTAxiomA}
\subsection{Geometry of the problem}
Let us consider a continuous-time mixing Axiom A dynamical system $\dot{x}=G(x)$ on a compact manifold $N\subset\mathbb{R}^d$,  where $x(t)=f^t(x_{in})$, with $x(t=0)=x_{in}\in N$ initial condition and $f^t$ evolution operator, is defined for all $t\in\mathbb{R}_{\geq 0}$. Let us define $\Omega$ as the attracting invariant set of the dynamical system , so that $\nu$ is the associated SRB measure with support $\Omega=\text{supp}(\nu)$. Let us now consider a smooth  observable $A$ whose maximum restricted to the support of  $\nu$ is unique, so that $\max(A)|_{\Omega}=A(x_0)=A_{max}$, $x_0\in \Omega$, and is, moreover, not a critical point, so that $\nabla A|_{x=x_0}\neq 0$, where the gradient is taken in $N$. Therefore, we have that the the  neutral manifold and the unstable manifold  are tangent to the manifold $A(x)=A_{max}$  in $x=x_0$. 
\begin{figure}[t!]
\includegraphics[width=0.81\columnwidth]{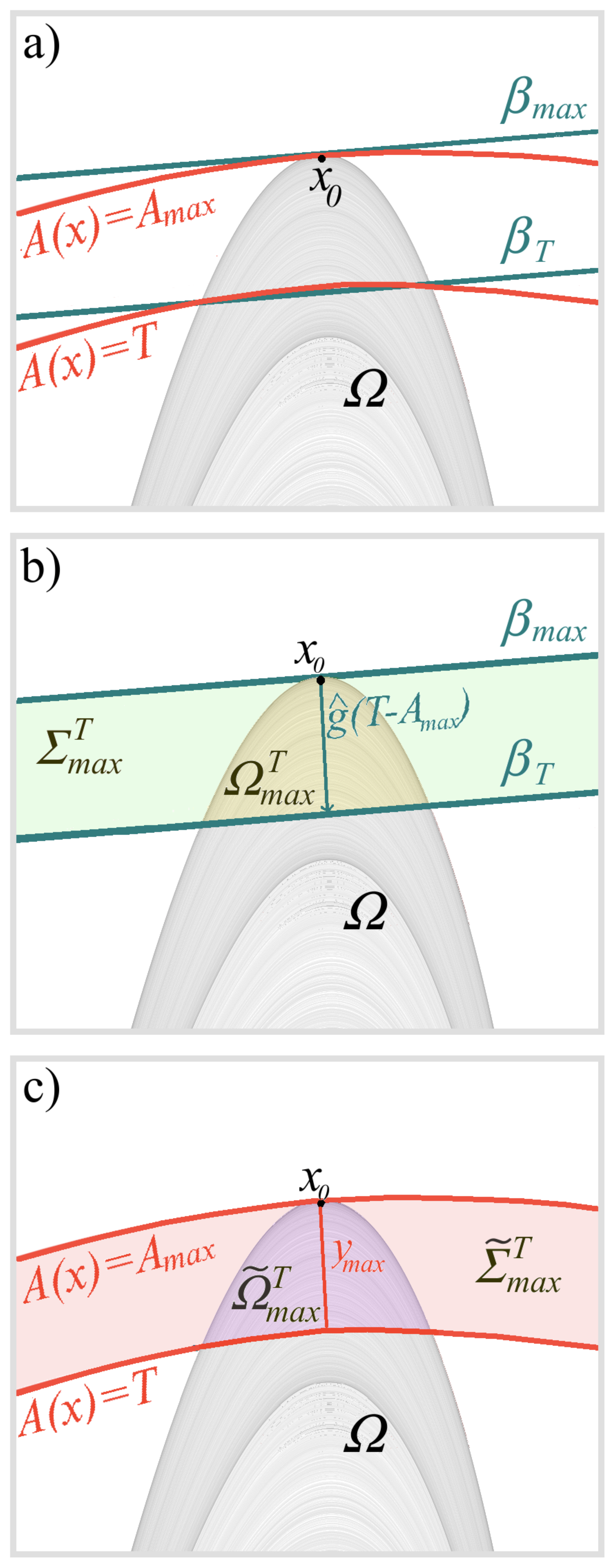}
\caption{A low-dimensional cartoon of the geometrical construction used for deriving the EVL for exceedances above the  threshold $T$ for the observable $A(x)$ such that $\max(A)|_{\Omega}=A_{max}$ is realized for $x=x_0$.  a) The manifolds $A(x)=A_{max}$ and $A(x)=T$ are depicted, together with the the attracting invariant set $\Omega$ and the two hyperplanes $\beta_{max}$ and $\beta_T$. $\beta_{max}$ is tangent to $A(x)=A_{max}$ in $x_0$ and $\beta_T$ is obtained from $\beta_{max}$ via translation along $(T-A_{max})\hat{g}$. b) The hyperplanes  $\beta_{max}$ and $\beta_T$ delimit the region $\Sigma^T_{max}$. Its intersection with $\Omega$ is $\Omega^T_{max}$ c) The manifolds $A(x)=A_{max}$ and $A(x)=T$ delimit the region $\tilde\Sigma^T_{max}$. Its intersection with $\Omega$ is $\tilde\Omega^T_{max}$. As $T\rightarrow A_{max}$, we have that $\Omega^T_{max}\rightarrow\tilde\Omega^T_{max}$.}
\label{fig1}
\end{figure}
We have that the intersection between the manifolds $A(x)=\tilde{A}$ and $\Omega$ is the empty set if $\tilde{A}>A_{max}$. We define as $\tilde{\Sigma}^T_{max} $  the subset of $\mathbb{R}^d$ included between the manifolds $A(x)=A_{max}$ and $A(x)=T$. Furthermore, we define as $\Sigma^T_{max} $  the subset of $\mathbb{R}^d$ included between the hyperplane $\beta_{max}$ tangent to the manifold $A(x)=A_{max}$ in $x=x_0$ and the hyperplane $\beta_{T}$, which is obtained by applying the translation given by the vector $(T-A_{max})\hat{g}$ to the hyperplane $\beta_{max}$, where $\hat{g}=\nabla A|_{x=x_0}/|\nabla A|_{x=x_0}|^2$. As $T\rightarrow A_{max}$, which is the limit of our interest, we have that $\tilde\Omega^T_{A_{max}}=\Omega \cap \tilde{\Sigma}^T_{A_{max}}$ and  $\Omega^T_{A_{max}}=\Omega \cap \Sigma^T_{A_{max}} $ become undistinguishable up to leading order. See Fig. \ref{fig1} for a a depiction of this geometrical construction. 

More in general, we denote as  ${\Omega}^U_V$, with $V>U>T$, the intersection between ${\Omega}$ and the subset of $\mathbb{R}^d$ included between the hyperplane $\beta_{U}$ and  $\beta_{V}$, where $\beta_{X}$, $X=U,V$ is obtained from $\beta_{max}$ by applying to it the translation given by the vector $(X-A_{max})\hat{g}$.

It is now clear that we observe an exceedance of the observable $A(x)$ above $T$ each time the systems visits a point belonging to ${\Omega}^T_{A_{max}} $. In more intuitive terms, and taking the linear approximation described above, an exceedance is realized each time the system visit a point $x\in\Omega$ whose distance $\text{dist}(x,\beta_{max})$ from the hyperplane $\beta_{max}$ is smaller than $y_{max}=(A_{max} -T)/|\nabla A|_{x=x_0}|$.  

We define the exceedances for the points $x\in{\Omega}^T_{A_{max}}$ as $z= A(x)-T$. An exceedance $z$ corresponds geometrically  to a distance $y=\text{dist}(x,\beta_{T})=z/|\nabla A|_{x=x_0}|$ from  $\beta_{T}$, and the maximum exceedance $A_{max}-T$ corresponds to a distance $(A_{max} -T)/|\nabla A|_{x=x_0}|$ between $x_0$ and $\beta_{T}$.

Therefore, $P(z>Z|z>0)=P(z>Z)/P(z>0)$. The probability $H_{T}(Z)$ of observing an exceedance of at least $Z$ given that an exceedance occurs is given by:
\begin{equation}
H_{T}(Z)\equiv\frac{\nu( \Omega^{T+Z}_{A_{max}} )}{\nu(  {\Omega}^T_{A_{max}}) }.
\label{measureball}
\end{equation}
where we have used the ergodicity of the system. Obviously, the value of the previous expression is 1 if $Z=0$. The expression contained in Eq. \eqref{measureball} monotonically decreases with $Z$ (as $\Omega^{T+Z_2}_{A_{max}} \subset \Omega^{T+Z_1}_{A_{max}} $ if $Z_1<Z_2$) and vanishes when $Z=A_{max}-T$. 

\subsection{Derivation of the Generalised Pareto Distribution Parameters for the Extremes of a Physical Observable}
We now wish to understand how to express the numerator and denominator of Eq. \ref{measureball} as a function of $T$, $Z$, and $A_{max}$.  We follow some of the ideas presented in \citet{holland2012} and use the fact that we are considering Axiom A systems. We define $D$ as the local dimension around $x_0$, such that
$$\lim_{r\rightarrow0}\frac{\log(\nu(B_r(x_0)))}{\log(r)}=D.$$
In order to proceed with the derivation of an extreme value law, such an asymptotics is not sufficient. In order to overcome some of the difficulties discussed in \citep{lucarinietal2012,faranda2011extreme}, one needs to assume that $$\nu(B_r(x_0)))\sim f_{x_0}(r)r^D,$$ where $f_{x_0}(r)$ is a slowly varying function of $r$ as $r\rightarrow 0$, possibly depending on $x_0$. We use the $\sim$ symbol as follows. We say that $f(x)\sim g(x)$ for $x\rightarrow y$ if $\lim_{x\rightarrow y} f(x)/g(x)=c$, $0<|c|<\infty$. Taking the assumption above, one can derive the extreme value laws for the distance observables discussed in \cite{lucarinietal2012}. Let's now estimate $\nu(  {\Omega}^T_{A_{max}})$ as a function of $y_{max}$ in the case of generic quadratic tangency between the the hyperplane $A(x)=A_{max}$  and the unstable manifold in $x=x_0$ 

In the case of  Axiom A systems, since the invariant measure is SRB, we have $D(x)=d_H$ almost everywhere on the attractor \citep{barr1999,galatolo2006}, where $d_H$ is the Hausdorff dimension. As discussed in \citep{barr1999}, we have that in this case all of the generalized Renyi dimensions have the same value. Moreover, we can conjecture that $D=d_{KY}$, where $d_{KY}$ is the Kaplan-Yorke dimension \citep{ruelle89}: 
\begin{equation}
d_{KY}=n+\frac{\sum_{k=1}^n \lambda_k}{|\lambda_{n+1}|}\label{dky},
\end{equation}
where the $\lambda_j$'s are the Lyapunov exponents of the systems, ordered from the largest to the smallest, $n$ is such that $\sum_{k=1}^n \lambda_k$ is positive and $\sum_{k=1}^{n+1} \lambda_k$ is negative. Following \citep{barr1999}, we can also write $d_{KY}=d_u+d_n+d_s$, where $d_s$, $d_u$ and $d_n$ are the dimensions  of the attractor $\Omega$ restricted to the stable, unstable and neutral directions, respectively, at the point $x=x_0$. We have that $d_u$ is equal to the number of positive Lyapunov exponents $\lambda_j^+$, $d_u=\#(\{\lambda_j>0\}, j=1,\ldots,d)$, $d_n$  for Axiom A systems is unitary, while $d_s$ is given by $d_s=d_{KY}-d_u-d_n$. Note that we can also express $d_{KY}$ as follows 
\begin{equation}
d_{KY}=d_u+d_n+[d_s]+\frac{\sum_{k=1}^n \lambda_k}{|\lambda_{n+1}|}=d_u+d_n+d_s,\label{dkyfin}
\end{equation}
so that $\{d_s\}=\sum_{k=1}^n \lambda_k/|\lambda_{n+1}| $, because the last term gives a positive contribution smaller than 1, and $d_u$ and $d_n$ are both integer.


We follow the construction proposed by \citet{holland2012} for low dimensional maps. We derive the result by considering the following heuristic argument. Near $x_0$, the attractor could be seen as the cartesian product of a multidimensional paraboloid (of dimension $d_u+d_n$ and of a fractal set of dimension $d_s$ immersed in $\mathbb{R}^{d-d_u-d_n}$. Note that this excludes for example conservative chaotic systems, whose attractor has the same dimension of the phase space, and systems that can be decomposed into a conservative part and a purely contractive part, whose attractor also has integer dimension. The mass of the paraboloid $\sim r^{d_u+d_n}$, where $r$ is the distance from the minimum. Instead, for each point of the paraboloid, the mass of corresponding fractal set $\sim h_{x_0}(l)l^{d_s}$, where $l$ is the distance along the the cartesian projection and $h_{x_0}(l)$ is a slowly varying function of $l$ as $l\rightarrow 0$, possibly depending on $x_0$. This is where the non-trivial slowly varying pre-factor is relevant appears.

In our case, $l =\gamma  y_{max}$ and $r=\kappa \sqrt{y_{max}}$: for the former relation we assume a generic relation between the direction of the gradient of $A$ and the stable directions, while the latter relation results from the functional form of the paraboloid, see also Fig. \ref{fig1}. Hence,  we obtain that $\nu(  {\Omega}^T_{A_{max}})\sim \tilde{h}_{x_0}(y_{max}) y_{max}^{\delta}$ where
\begin{equation}
\delta=d_s+(d_u+d_n)/2.
\label{deltadef}
\end{equation}
and $\tilde{h}_{x_0}(y_{max})={h}_{x_0}(\gamma y_{max})$ is also a slowly varying function of its argument. This construction can be made more formal by considering the disintegration of the SRB measure $\nu$ along the stable and unstable directions of the flow \citep{ruelle89}.  

As a side note, we emphasize that in the case of more general tangencies between the the unstable and neutral manifold and the  manifold $A(x)=A_{max}$ , we have that  $\delta=d_s+\sum_{j=1} d_u(2j)/(2j)+d_n(2j)/(2j)$, where $d_u(2j)$ ($d_n(2j)$ ) gives the number of directions along the unstable (neutral) manifold where a tangency of order $2j$ with the manifold $A(x)=A_{max}$ is found. We obviously have that $\sum_{j=1} d_u(2j)=d_u$.  

We continue our discussion considering the case of generic tangency.  Following the same argument as above, we have that $\nu(  {\Omega}^{T+Z}_{A_{max}})\sim \tilde{h}_{x_0}(y_{max}-y)(y_{max}-y)^{\delta}$, where $y=Z//|\nabla A_{x=x_0}|$. We define $$\alpha=1-y/y_{max}=1-Z/(A_{max}-T))$$ and obtain $\nu(  {\Omega}^{T+Z}_{A_{max}})\sim {\tilde{h}}_{x_0}(\alpha y_{max})\alpha^\delta (y_{max}-y)^{\delta}$.
Using the definition of slowly varying function and considering Eq. \ref{measureball}, we derive that in the limit $T\rightarrow A_{max}$: 
\begin{equation}
H_{T}(Z)=\alpha^\delta=\left(1-\frac{Z}{A_{max}-T}\right)^\delta.
\label{measureball2}
\end{equation}
Note that the corresponding cdf is given by $F_{T}(Z)=1-H_{T}(Z)$. Comparing Eqs. \ref{GPD} and Eqs. \ref{measureball2}, one obtains that $F_{T}(Z)$ belongs to the GPD family, and that the GPD parameters can be expressed as follows: 
\begin{align} 
&\xi=-1/ \delta \label{xires}\\ 
&\sigma=(A_{max}-T)/\delta. \label{sigmares}
\end{align}
These results complement and extend what obtained in \citep{holland2012} using the GEV approach. 

It is important to remark that Eq. \ref{measureball2} has been obtained  in the limit of $T\rightarrow A_{max}$, and under the assumption that $\nu(B_{r}(x_0)$ ia a regularly varying function of degree D as $r\rightarrow0$. When considering a finite range $A_{max}-T$, one should not expect deviations of the empirical distributions of extremes of $A$ from what prescribed in Eq. \ref{measureball2}, which are intrinsic to the fractal nature of the measure. See also discussion and Fig. 1 in \cite{lucarinietal2012}. When finite ranges for $A$ are considered, one expects that, in some averaged sense,  Eq. \ref{measureball2} fits well the distributions of extremes of $A$ and  Eqs. \ref{xires} and \ref{sigmares} give the value of the two relevant parameters of the GPD, analogously to the idea that the number of points of the attractor at distance smaller than a small but finite $r$ from the point $x_o$ scales approximately, on the average as $r^D$. 

\subsection{Comments}
Equation \ref{xires} provides a very valuable information,as it shows that the shape parameter $\xi$ of the GPD does not depend on the considered observable, but only on the dimensions of the stable and of the unstable manifold. Moreover, the shape parameter is always negative, which is hardly surprising as we are considering  compacts attractor and a well-behaved observable, whose values on the attractor have an upper bound.  Note that for Axiom A systems, $d_s$ and $d_u$ are constant almost everywhere in the attractor $\Omega$, so that the information gathered for $x=x_0$ has a global value. Therefore, the expression for $\xi$ is universal, in the sense that we can gather fundamental properties of the dynamical system by looking at the shape parameters of the extremes of a generic observables with the properties described above. Note also that $\delta$ can be used to put upper and lower bounds to the Kaplan-Yorke dimension of the system, as $d_{KY}=d_s +d_u+d_n>d_s +(d_u+d_n)/2= \delta=-1/ \xi$ and $d_{KY}=d_s +d_u+d_n<2d_s +d_u+d_n= 2\delta=-2/ \xi$, so that $-1/\xi<d_{KY}<-2/\xi$. On the other hand, these inequalities can be read as constraints to the shape parameters of the extremes of a general observable for a system for which we know the Kaplan-Yorke dimension:  $d_{KY}/2<-1/\xi<d_{KY}$.


On the other hand, the expression we obtain for $ \sigma$ provides clear support for calling it the scale parameter. We derive, as anticipated, that $ \sigma>0$ and we observe that it is proportional to the actual range of values considered as extremes of the observable $A(x)$, by incorporating the difference between the absolute maximum of the observable $A_{max}$ and the selected threshold $T$. Therefore, if we consider as observable $A_1(x)= \alpha A(x)$, with $\alpha>0$ and take as threshold for $A_1(x)$ the value $\alpha T$, we have that $\xi_{A_1}=\xi_A$ and $\sigma_{A_1}=\alpha\sigma_A$. In physical terms, $\sigma$ changes if we change the unit of measure of the observable, whereas $\xi$ does not.  More generally, we can make the following construction. Let's define  $\min(A)|_{\Omega}=A_{min}$. If we select an observable $A_2(x)=\Phi(A(x))$, with $\max(\Phi)|_{[A_{min},A_{max}]}=\Phi(A_{max})$, $\Phi$ differentiable and $d\Phi(y)/dy$ positive in a sufficiently wide neighbourhood around $y=A(x_{max})$ so to ensure monotonicity of $A_2(x)$ near $x=x_{max}$, we get $\xi_{A_2}=\xi_A$ and  $\sigma_{A_2}=\gamma\sigma_A$, where $\gamma=d\Phi(y)/dy|_{y=A(x_{max})}$. 
 
\subsection{From the Extremes to the Partial Dimensions along the Stable and Unstable Directions of the Flow}
It is worth considering the following strategy of investigation of the local properties of the invariant measure near $x=x_{0}$, where $A(x_0)=A_{max}$. By performing statistical inference of the extremes of $A$ we can deduce as a result of the data fitting the best estimate of $\xi_A=1/\delta$. If, following \citep{lucarinietal2012}, we select as observable, e.g. $B(x)=C-(\text{dist}(x,x_0))^{\beta} $, $\beta>0$, we have that the extremes of the observable $B$ feature as shape parameter  $\xi_B=-\beta/D=-\beta/d_{KY}$ and scale parameter $\sigma_B=(C-\tau)\beta/D=(C-\tau)\beta/d_{KY}$ \citep{lucarinietal2012}, where $C$ is a constant and $\tau$ is the chosen threshold. 

We can then easily derive:
\begin{align}
\frac{2}{\xi_A}-\frac{2\beta}{\xi_B}&=d_u+d_n \label{du}\\
\frac{\beta}{\xi_B}-\frac{2}{\xi_A}&=d_s \label{ds}
\end{align}
where, as discussed above, we can take $d_n=1$. Therefore, using rather general classes of observables, we are able to deduce the partial dimensions along the stable and unstable manifolds, just by looking at the properties of extremes related to $x=x_0$. It is to be noted that, as clear from the results presented in \citep{lucarinietal2012}, similar conclusions can be drawn choosing powers of $\text{dist}(x,x_0)$ forms for $B$ are possible. Note that, more generally, $d_u$ and $d_s$ can be deduced from the knowledge of any pair of values $(\xi_A,\xi_B)$, $(\sigma_A,\xi_B)$, $(\xi_A,\sigma_B)$, and $(\sigma_A,\sigma_B)$.

\subsection{Expressing the shape parameter in terms of the GPD moments and of the invariant measure of the system}
We consider the physical observable $A$. 
 We denote by 
\begin{align}
f_{GPD}(z;\xi_A,\sigma_A) & = \frac{\text{d}}{\text{d}z}\left(F_{GPD}(z;\xi_A,\sigma_A)\right)\nonumber \\& = \frac{1}{\sigma_A}\left(1+ \frac{\xi_A z}{\sigma_A}\right)^{-1/\xi_A-1}
\end{align}
 the density corresponding to the cumulative distribution given in Eq. \ref{GPD}. We can express its first two moments as follows:
\begin{align}
\int_0^{-\sigma_A/\xi_A}dz  \hspace{1mm} z\hspace{1mm}  f_{GPD}(z;\xi_A,\sigma_A)&=\frac{\sigma_A}{1-\xi_A}=\mu_1\\
\int_0^{-\sigma_A/\xi_A}dz  \hspace{1mm} z^2\hspace{1mm}   f_{GPD}(z;\xi_A,\sigma_A)&=\frac{2\sigma_A^2}{(1-\xi_A)(1-2\xi_A)}=\mu_2.
\end{align} 
It is easy to derive that 
\begin{equation}
\xi_A=\frac{1}{2}\left(1-\frac{\mu_1^2}{\mu_2-\mu_1^2}\right)=\frac{1}{2}\left(1-\frac{1}{id_A}\right)\label{id}
\end{equation}
and
\begin{equation}
\sigma_A=\frac{\mu_1\mu_2}{2(\mu_2-\mu_1^2)}
\end{equation}
where we indicate explicitly that we refer to the observable $A$ and 
we have introduced the index of dispersion $id_A$, the ratio between the variance and the squared first moment of the considered stochastic variable.


We now try to connect the previous formulas to the properties of the invariant measure of the dynamical system. As we know, the GPD is the exact asymptotic model for the extremes of the observable $A$, so that we can express the results in terms of the conditional invariant measure as follows:
\begin{align}
\mu_1^{T }&=\frac{\int \nu(dx)\Theta(A(x)-T)(A(x)-T)}{\int \nu(dx)\Theta(A(x)-T)}=\frac{\langle \tilde{A}_1 ^{T}\rangle }{\langle \tilde{A}_0 ^{T}\rangle}\label{obse1}\\
\mu_2^{T }&=\frac{\int \nu(dx)\Theta(A(x)-T)(A(x)-T)^2}{\int \nu(dx)\Theta(A(x)-T)}=\frac{\langle \tilde{A}_2 ^{T}\rangle }{\langle \tilde{A}_0 ^{T}\rangle}\label{obse2}.
\end{align} 
where $\Theta$ is the usual Heaviside distribution and, in general, 
\begin{equation}\langle \tilde{A}_n ^{T}\rangle = \int \nu(dx)\Theta(A(x)-T)(A(x)-T)^n.\label{Agen}\end{equation}
We then obtain the following expression for the shape and dispersion parameters, respectively:
\begin{equation}
\xi_A^{T }=\frac{1}{2}\left(1-\frac{(\langle \tilde{A}_1 ^{T}\rangle)^2} {\langle \tilde{A}_2 ^{T}\rangle \langle \tilde{A}_0 ^{T}\rangle -(\langle \tilde{A}_1 ^{T}\rangle)^2 }\right)\label{ruellexi},
\end{equation}
and
\begin{equation}
\sigma_A^{T }=\frac{1}{2}\frac{\langle \tilde{A}_1 ^{T}\rangle \langle \tilde{A}_2 ^{T}\rangle}{ \langle \tilde{A}_2 ^{T}\rangle \langle \tilde{A}_0 ^{T}\rangle -\langle \tilde{A}_1 ^{T}\rangle^2}\label{ruellesigma},
\end{equation}
where these result are exact in the limit for $T\rightarrow A_{max}$. 
As a check, it is useful to verify that the right hand side of Eq. \ref{ruellexi} gives the same general result as given in Eq. \ref{xires}. By definition we have:
\begin{align}
\nu(  {\Omega}^T_{A_{max}})&=\langle \tilde{A}_0 ^{T}\rangle = \int \nu(dx)\Theta(A(x)-T)\nonumber\\
&\sim \tilde{h}_{x_0}(A_{max}-T)(A_{max}-T)^{\delta^\epsilon}.
\end{align}
We derive the following expression using repeatedly the distributional relation $x$ $d/dx [\Theta(x)]=0$:
\begin{align}
\langle \tilde{A}_0 ^{T}\rangle &=-\frac{\text{d}}{\text{d}T}\langle \tilde{A}_1 ^{T}\rangle  \label{a0}\\
\langle \tilde{A}_1 ^{T}\rangle &=\frac{1}{2}\frac{\text{d}}{\text{d}T}\langle \tilde{A}_2 ^{T}\rangle \label{a1}
\end{align} 
so that
\begin{equation}
\langle \tilde{A}_1 ^{T}\rangle\sim\frac{\tilde{h}_{x_0}(A_{max}-T)}{(\delta+1)} (A_{max}-T)^{\delta+1} \label{a1b}
\end{equation}
and \begin{equation}
\langle \tilde{A}_2 ^{T}\rangle\sim\frac{2\tilde{h}_{x_0}(A_{max}-T)}{(\delta+1)(\delta+2)} (A_{max}-T)^{\delta+2}\label{a2b}.
\end{equation}
By plugging these expression into  Eq. \ref{ruellexi}, we indeed obtain $\xi=-1/\delta$, which agrees with Eq. \ref{xires}. 

We also note that it is possible to generalize the results given in Eqs. \ref{a0}-\ref{a2b}. Using the fundamental theorem of calculus, it is possible to derive that:  
$$\langle \tilde{A}_n ^{T}\rangle = \int_T^{A_{max}} dz \hspace{1mm} n(z-T)^{n-1}\langle \tilde{A}_0^{z } \rangle$$. 


Moreover, it is notable that what presented in this subsection can be replicated step by step for the distance observables observable $B(x,x_0)=C-\text{dist}(x,x_0)^{\beta}$ discussed above. We obtain:
\begin{equation}
\xi_B^{T }=\frac{1}{2}\left(1-\frac{(\langle \tilde{B}_1 ^{T}\rangle)^2} {\langle \tilde{B}_2 ^{T}\rangle \langle \tilde{B}_0 ^{T}\rangle -(\langle \tilde{B}_1 ^{T}\rangle)^2 }\right)\label{ruellexibis}
\end{equation}
and
\begin{equation}
\sigma_B^{T }=\frac{1}{2}\frac{\langle \tilde{B}_1 ^{T}\rangle \langle \tilde{B}_2 ^{T}\rangle}{ \langle \tilde{B}_2 ^{T}\rangle \langle \tilde{B}_0 ^{T}\rangle -\langle \tilde{B}_1 ^{T}\rangle^2}\label{ruellesigmabis},
\end{equation}
where the quantities $\langle \tilde{B}_j ^{T}\rangle$, $j=0, 1, 2$ are constructed analogously to how described in Eq. \ref{Agen}. 

We wish to remark that Eqs. \ref{ruellexi}-\ref{ruellesigma} and Eqs. \ref{ruellexibis}-\ref{ruellesigmabis} could in fact provide a very viable method for estimating the GPD parameters from data, since moments estimator are in general more stable than maximal likelihood methods, and then deriving the value of $d_u$ and $d_s$ using Eqs. \ref{du}-\ref{ds}.

\section{Response Theory for the Extremes of General Observables}
\label{ResponseEVT}

We wish to present some ideas on how to use response theory and the specific expressions given in Eqs \ref{xires}-\ref{sigmares} to derive a response theory for extremes of physical and distance observables in Axiom A dynamical systems.  Let's assume that we alter the Axiom A dynamical system under consideration as $\dot{x}=G(x)\rightarrow \dot{x}=G(x)+\epsilon X(x)$, where $\epsilon$ is a small parameter and $X(x)$ is a smooth vector field, so that the evolution operator is transformed as $f^t\rightarrow{f}_\epsilon^t$ and the invariant measure is transformed as $\nu\rightarrow \nu_\epsilon$.  Ruelle's response theory allows to express the change in the expectation value of a general measurable observable $\Psi(x)$ as a perturbative series as $\langle \Psi \rangle^\epsilon=\langle \Psi \rangle_0+\sum_{j=1}^\infty \epsilon^j \langle \Psi^{(j)} \rangle_0$, with $j$ indicating the order of perturbative expansion, where $$\langle \Psi \rangle^\epsilon=\int\nu_{\epsilon}(dx) \Psi(x)$$ is the expectation value of $\Psi$ over the perturbed invariant measure and $$\langle \Psi \rangle_0=\int\nu(dx) \Psi(x)$$ defines the unperturbed expectation value of $\Psi$. The term corresponding to the perturbative order of expansion $j$ is given by $\langle \Psi^{(j)} \rangle_0$, where $\Psi^{(j)}$ can be expressed in terms of the time-integral of a suitably defined Green function \cite{ruelle98}. At this stage, we limit ourselves to the linear response of the system. 
We consider the following useful formula: $$\frac{\text{d}^n\langle \Psi \rangle^\epsilon}{\text{d}\epsilon^n}\bigg|_{\epsilon=0}=n!\langle \Psi^{(n)} \rangle_0.$$ 
and take into account the $n=1$ case. 

\subsection{Sensitivity of shape parameter as determined by the changes in the moments of the distribution}

We wish to propose a linear response formula for the parameter $\xi_A$ using Eqs. \ref{obse1}-\ref{ruellexi}. We start by considering that in Eq. \ref{ruellexi} the shape parameter is expressed for every $T<A_{max}$ as a function of actual observables of the system. Unfortunately, in order to apply Ruelle's response theory, we need the observables to be smooth, which is in contrast with the presence of the $\Theta$ in the definition of the terms $\langle \tilde{A}_j ^{T}\rangle^\epsilon$. Nonetheless, replacing the $\Theta$'s with a smooth approximation $\Theta_S$ , the Ruelle response theory can be rigorously applied. We now consider a sequence of approximating $\Theta_S^m$ such that the measure of the support of $\Theta-\Theta_S^m$ is smaller than $\delta_m=(A_{max}-T)/m$. It is reasonable to expect that as $\delta_m\rightarrow0$, the effect of the smoothing becomes negligible, because a smaller and smaller portion of the extremes is affected, and the response of the smoothed observable approaches that of  $\langle \tilde{A}_j ^{T}\rangle^\epsilon$. Therefore, we can retain the $\Theta$ in the definition of the $\langle \tilde{A}_j ^{T}\rangle^\epsilon$ and define rigorously for every $T<A_{max}$: 
 \begin{equation}
\frac{\text{d} \xi_A^{T,\epsilon}}{\text{d} \epsilon}\bigg|_{\epsilon=0}=-\frac{1}{2}\frac{\text{d} }{\text{d} \epsilon}\left\{\frac{(\langle \tilde{A}_1 ^{T}\rangle^\epsilon)^2} {\langle \tilde{A}_2 ^{T}\rangle^\epsilon \langle \tilde{A}_0 ^{T}\rangle^\epsilon -(\langle \tilde{A}_1 ^{T}\rangle^\epsilon)^2 }\right\}\bigg|_{\epsilon=0}\label{sensixi2}
\end{equation}
and
\begin{equation}
\frac{\text{d} \sigma_A^{T,\epsilon}}{\text{d} \epsilon}\bigg|_{\epsilon=0}=\frac{1}{2}\frac{\text{d} }{\text{d} \epsilon}\left\{\frac{\langle \tilde{A}_1 ^{T}\rangle^\epsilon \langle \tilde{A}_2 ^{T}\rangle^\epsilon}{ \langle \tilde{A}_2 ^{T}\rangle^\epsilon \langle \tilde{A}_0 ^{T}\rangle^\epsilon -(\langle \tilde{A}_1 ^{T}\rangle^\epsilon)^2}\right\}\bigg|_{\epsilon=0}\label{sensisigma2},
\end{equation}

By expanding the derivative in Eq. \ref{sensixi2}, the previous expression can be decomposed in various contributions entailing the linear response of the system to the $\epsilon$ perturbation for the observables $ \langle \tilde{A}_0 ^{T}\rangle^\epsilon$, $ \langle \tilde{A}_1 ^{T}\rangle^\epsilon $, $ \langle \tilde{A}_2 ^{T}\rangle^\epsilon $ and their values in the unperturbed case for $\epsilon=0$.  

%
%

We wish to remark the special relevance of the observable $\langle \tilde{A}_0 ^{T}\rangle^\epsilon$, which is normalizing factor in Eqs. \ref{obse1}-\ref{obse2}, and, in practice, measures the fraction of above-$T$-threshold events. Therefore, once $T$ is chosen, the sensitivity  of $\langle \tilde{A}_0 ^{T}\rangle^\epsilon$ with respect to $\epsilon$ informs on whether the $\epsilon$-perturbation to the vector flow leads to an increase or decrease in the number of extremes. We obtain:
\begin{align}
\frac{\text{d}\langle A_0^T \rangle^\epsilon}{\text{d}\epsilon}\bigg|_{\epsilon=0}& =\int d\tau \left \langle  X_k(x)\partial_k \Theta\left(A(x(\tau)-T\right)\right\rangle_0 \nonumber \\
& =\int d\tau \left\langle X_k(x) \partial_k A(x(\tau) \delta\left(A(x(\tau)-T\right)\right\rangle_0\nonumber\\
&=\int d\tau \langle X_k(x) \partial_k  x_i(\tau) \partial_{x_i(t)} A(x(t))\nonumber \\  
& \quad \quad \quad \times  \delta\left(A(x(\tau)-T\right)\rangle_0\nonumber  \\
\end{align}
where $\delta$ is the derivative of the $\Theta$ function with all the caveats discussed above. The formula can be interpreted as follows. In the last formula, $\partial_k  x_i(\tau) $ is the adjoint of the tangent linear of the unperturbed flow, and $\partial_{x_i(t)}$ indicates the partial derivative with respect to the variable $x_i(t)$.  At each instant $\tau$ we consider, in the unperturbed system, all the trajectories starting in the infinite past from points distributed according to the invariant measure such that the observable $A$ has value equal to $T$. For each of these trajectories, we can measure whether the presence of the perturbation field $X(x)$ would lead to a decrease or increase in $A$ at time $\tau$. Summing over all trajectories, we get whether there is a net positive or negative change in the above threshold events at time $\tau$. We integrate over $\tau$ and get the final result. 
Considering the geometrical construction given in Fig. \ref{fig1}, the previous formula can also be approximated as follows:
\begin{align}
\frac{\text{d}\langle A_0^T \rangle^\epsilon}{\text{d}\epsilon} \bigg|_{\epsilon=0}&\approx \int d\tau \langle X_k(x) \partial_k  x_i(\tau) \partial_i A|_{x=x_0}\nonumber \\ & \quad \quad \quad \times   \delta\left(A(x(\tau)-T\right)\rangle_0 
\end{align}

Therefore, Eqs. \ref{sensixi2}-\ref{sensisigma2} provides recipes for computing the sensitivity of $\xi_A^{T,\epsilon}$ and $\sigma_A^{T,\epsilon}$ at $\epsilon=0$ for any case of practical interest, where $A_{max}-T$ is indeed finite, because in order to collect experimental data or process the output of numerical simulations we need to select a threshold which is high enough for discriminating true extremes and low enough for allowing a sufficient number of samples to be collected for robust data processing. Note that all statistical procedures used in estimating GPD parameters from data are actually based on finding a reasonable value for $T$ such that both conditions described above apply by testing that parameters' estimates do not vary appreciably when changing $T$ \citep{pickands1975statistical}. So, if when investigating the extremes of $A$ for the unperturbed and $\epsilon-$perturbed dynamics we find a common value of $T$ such that the GPD statistical inference of extremes is satisfactory, Eqs. \ref{sensixi2}-\ref{sensisigma2} provide correct formulas for the sensitivities..

We wish to underline that apparently formal problems emerge when taking the limit in Eqs.\ref{sensixi2}-\ref{sensisigma2} for higher and higher values of $T$. It is indeed not clear at this stage whether 
\begin{equation}\lim_{T\rightarrow A_{max}} \frac{\text{d} \xi_A^{T,\epsilon}}{\text{d} \epsilon}\bigg|_{\epsilon=0} =\lim_{T\rightarrow A_{max}}\lim_{\epsilon\rightarrow 0} \frac{\xi_A^{T,\epsilon}-\xi_A^{T,0}}{\epsilon}\label{limit1}
\end{equation}
exists, because we cannot apply the smoothing argument presented above in the limit of vanishing $A_{max}-T$. Moreover, it is not clear whether such limit is equal to 
\begin{equation}
\lim_{\epsilon\rightarrow 0} \lim_{T\rightarrow A_{max}}\frac{\xi_A^{T,\epsilon}-\xi_A^{T,0}}{\epsilon},\label{limit2}
\end{equation}
which seems at least as well suited for describing  the change of the shape observable given in Eq. \ref{ruellexi} due to an $\epsilon-$perturbation in the dynamics. Obviously, if the two limits given in Eqs. \ref{limit1} and \ref{limit2} exist and are equal, then a rigorous response theory for $\xi_A$ can be established. Same applies when considering the properties of $\sigma_A^{T,\epsilon}$.



The same derivation and discussion can be repeated for the $B$ observables introduced above and we can derive the corresponding formulas for $\text{d} \xi_B^{T,\epsilon}/\text{d} \epsilon|_{\epsilon=0}$ and $\text{d} \sigma_B^{T,\epsilon}/\text{d} \epsilon|_{\epsilon=0}$, 
%
where the relevant limit for $T$ is $T\rightarrow C$.

Let's try to give a more intuitive interpretation to the results given above. Let's consider Eq. \ref{id} and assume that, indeed, $\xi$ is differentiable with respect to $\epsilon$. We have:
\begin{equation}
\frac{\text{d} \xi_A^\epsilon}{\text{d} \epsilon}\bigg|_{\epsilon=0}=-\frac{1}{2}\frac{\text{d}}{\text{d} \epsilon} \left\{\frac{1}{id_A^\epsilon}\right\}\bigg|_{\epsilon=0}=\frac{1}{2 id{_A^{\epsilon} }^2}\frac{\text{d}}{\text{d} \epsilon} \left\{id_{A}^{\epsilon}\right\}\bigg|_{\epsilon=0}.\label{sensixib}
\end{equation}
which implies that the sensitivity of the shape parameter is half of the opposite of the sensitivity of the inverse of the index of dispersion $id_A$. Therefore, a positive sensitivity of the index of dispersion (larger relative variability of the extremes of the observable $A$ with positive values of $\epsilon)$ implies a larger value (closer to 0) of $\xi_A$, and so the possibility that larger and larger extremes are realized. Same interpretation applies for the $B$ observables.  

\subsection{Sensitivity of the shape parameter as determined by the modification of the geometry}
In the previous subsection we have shown that the Ruelle response theory supports the idea that the shape parameters descriptive of the extremes of both the physical observables $A$ and the distance observables $B$ change with a certain degree of regularity when considering $\epsilon-$perturbations to the dynamics. 

In this subsection, we wish  to look at the sensitivity of extremes with respect to perturbation from another angle, \textit{i.e.} through the relationship between the shape parameters $\xi_A$ and $\xi_B$ and the partial dimension of the attractor along the stable, neutral and unstable manifolds of the underlying dynamical system, see Eqs. \ref{du}-\ref{ds}. As long as the $\epsilon$-perturbation is small, the modified dynamical system belongs to the Axiom A family, so that the results presented above apply. Therefore, we can write in more general terms: 
\begin{align}
\xi_A^\epsilon=-1/\delta^\epsilon &=-1/(d_s^\epsilon+d_u^\epsilon/2+d_n^\epsilon/2) \label{xiadyn}\\
\xi_B^\epsilon=-\beta/d_{KY}^\epsilon& =-\beta/(d_s^\epsilon+d_u^\epsilon+d_n^\epsilon)\label{xibdyn}.  
\end{align}
In the following, we introduce somewhat carelessly derivatives with respect to $\epsilon$ of quantities that are not, \textit{a priori}, differentiable. The main point we want to make is that if $\xi_A$ and $\xi_B$ are differentiable with respect to $\epsilon$, then various quantities describing the structure of the attractor are also differentiable. Therefore, the existence of the limits given in Eqs. \ref{limit1} and \ref{limit2} (and their equivalent for the $B$ observables) would have far-reaching consequences. We will discuss the obtained results at the end of the calculations. Another \textit{caveat} we need to mention is that Eqs. \ref{xiadyn}-\ref{xibdyn} are in general true almost anywhere, so that we may have to interpret the derivatives in this section in some suitable weak form.  

It seems relevant to add the additional hypothesis of strong transversality for the unperturbed flow, which is equivalent to invoking structural stability \citep{ruelle89}. We take such pragmatic point of view and proceed assuming that derivatives with respect to $\epsilon$ are well defined. Linearizing the dependence of $\xi_A$ on $\epsilon$ around $\epsilon=0$ in Eq. \ref{xiadyn}, we obtain:
\begin{align}
\frac{\text{d} \xi_A^\epsilon}{\text{d}\epsilon}\bigg|_{\epsilon=0}=\left\{\frac{\text{d}\xi_A^\epsilon}{\text{d} (d_s^\epsilon)}\frac{\text{d}( d_s^\epsilon)}{\text{d} \epsilon}\right\}\bigg|_{\epsilon=0}&+\left\{\frac{\text{d}\xi_A^\epsilon}{\text{d} (d_u^\epsilon)}\frac{\text{d} (d_u^\epsilon)}{\text{d} \epsilon}\right\}\bigg|_{\epsilon=0} \nonumber \\  & +\left\{\frac{\text{d}\xi_A^\epsilon}{\text{d} (d_n^\epsilon)}\frac{\text{d} (d_n^\epsilon)}{\text{d} \epsilon}\right\}\bigg|_{\epsilon=0}.
\end{align}
We have that $\text{d}( d_u^\epsilon)/\text{d} \epsilon|_{\epsilon=0}=\text{d}( d_n^\epsilon)/\text{d} \epsilon|_{\epsilon=0}=0$, as, thanks to structural stability, small perturbations do not alter the qualitative properties of the dynamics, and cannot change in a step-wise way the number of expanding or neutral directions. We now separate the quantity $d_s$ into its integer component and the rest, which is generically non vanishing: 
\begin{equation}
\frac{\text{d} \xi_A^\epsilon}{\text{d}\epsilon}\bigg|_{\epsilon=0}=\left\{\frac{\text{d}\xi_A^\epsilon}{\text{d} (d_s^\epsilon)}\frac{\text{d}( [d_s^\epsilon])}{\text{d} \epsilon}\right\}\bigg|_{\epsilon=0}+\left\{\frac{\text{d}\xi_A^\epsilon}{\text{d}(d_s^\epsilon)}\frac{\text{d}( \{d_s^\epsilon\})}{\text{d} \epsilon}\right\}\bigg|_{\epsilon=0}.
\end{equation}
Only the last term is different from zero, because, thanks to structural stability, $\text{d}( [d_u^\epsilon])/\text{d} \epsilon|_{\epsilon=0}=0$. Using Eq. \ref{dkyfin}, we obtain:
\begin{equation}
\frac{\text{d} \xi_A^\epsilon}{\text{d}\epsilon}\bigg|_{\epsilon=0}=\left\{\frac{\text{d}\xi_A^\epsilon}{\text{d} (d_s^\epsilon)}\frac{\text{d} \frac{\sum_{k=1}^n \lambda^\epsilon_k}{|\lambda^\epsilon_{n+1}| }}{\text{d} \epsilon}\right\}\bigg|_{\epsilon=0}\label{dxiky1}
\end{equation}
where $n$ is defined as in Eq. \ref{dky}. Since $n$ is not altered for infinitesimal $\epsilon$ perturbations to the dynamical system, we have that, using Eq. \ref{dky}:
\begin{equation}
\frac{\text{d} \frac{\sum_{k=1}^n \lambda^\epsilon_k}{|\lambda^\epsilon_{n+1} |}}{\text{d} \epsilon}\bigg|_{\epsilon=0}=\frac{\text{d} (d^\epsilon_{KY})}{\text{d}\epsilon}\bigg|_{\epsilon=0}.\label{dxiky2}
\end{equation}
Expanding the previous expressions, the final formula reads as follows:
\begin{equation}
\frac{\text{d} \xi_A^\epsilon}{\text{d}\epsilon}\bigg|_{\epsilon=0}=\left\{\frac{1}{(d^\epsilon_s+d^\epsilon_u/2+d^\epsilon_n/2)^2}\frac{\text{d} (d^\epsilon_{KY})}{\text{d}\epsilon}\right\}\bigg|_{\epsilon=0}.\label{dxiky3}
\end{equation} 
This implies that the shape parameter $\xi$ increases, thus attaining a value closer to zero ($\xi_A$ is always negative) when the perturbation increases the Kaplan-Yorke dimension of the attractor, so, in qualitative sense, if it favors "forcing" over "dissipation".  This matches quite well, at least qualitatively, with the discussion following Eq. \ref{sensixib}.

We have that, when considering a distance observable of the form $B(x)=-\text{dist}(x,x_0)^{\beta}$, following the same steps described above one gets the following  result:
\begin{equation}
\frac{\text{d} \xi_B^\epsilon}{\text{d}\epsilon}\bigg|_{\epsilon=0}=\left\{\frac{\beta}{d{^\epsilon_{KY}}^{2}}\frac{\text{d} (d^\epsilon_{KY})}{\text{d}\epsilon}\right\}\bigg|_{\epsilon=0}\label{dxiky3bis};
\end{equation}
such result can be easily generalized by considering the class of observables described in \cite{lucarinietal2012}.

Combining Eq.  \ref{sensixi2} with Eq. \ref{dxiky3}, and the derivative with respect to $\epsilon$ of Eq. \ref{ruellexibis} with Eq. \ref{dxiky3bis}, we can derive two expressions for the derivative of the sensitivity of the Kaplan Yorke dimension at $\epsilon=0$:
\begin{align}
\frac{\text{d} (d^\epsilon_{KY})}{\text{d}\epsilon}\bigg|_{\epsilon=0}&=-\left\{\frac{(d^\epsilon_s+d^\epsilon_u/2+d^\epsilon_n/2)^2}{2}\right\}\bigg|_{\epsilon=0}\times \nonumber\\&
\times \left\{\frac{\text{d} }{\text{d} \epsilon}\frac{(\langle \tilde{A}_1 ^{T}\rangle^\epsilon)^2} {\langle \tilde{A}_2 ^{T}\rangle^\epsilon \langle \tilde{A}_0 ^{T}\rangle^\epsilon -(\langle \tilde{A}_1 ^{T}\rangle^\epsilon)^2 }\right\}\bigg|_{\epsilon=0}\label{ddky1} \\
&=-\left\{\frac{d^\epsilon{_{KY}}^2}{2\beta}\right\}\bigg|_{\epsilon=0}\times \nonumber \\& \times \left\{\frac{\text{d} }{\text{d} \epsilon}\frac{(\langle \tilde{B}_1 ^{T}\rangle^\epsilon)^2} {\langle \tilde{B}_2 ^{T}\rangle^\epsilon \langle \tilde{B}_0 ^{T}\rangle^\epsilon -(\langle \tilde{B}_1 ^{T}\rangle^\epsilon)^2 }\right\}\bigg|_{\epsilon=0}\label{ddky2}
\end{align}
where we take the limit for $T\rightarrow A_{max}$ in Eq. \ref{ddky1} and $T\rightarrow 0$ in Eq. \ref{ddky2}.

The previous results imply that if one of $\xi_A$, $\xi_B$ or the Kaplan-Yorke dimension of the underlying Axiom A system change smoothly with $\epsilon-$perturbations to the dynamics, so do the other two quantities. This may suggest ways to study the regularity of the Kaplan-Yorke dimension by resorting to the analysis of the regularity of a much simpler expressions involving moments of given observables. 

This result provides useful insight also not considering the problematic limits discussed above. Taking a more qualitative point of view, this suggests that when considering small perturbation in the dynamics  of chaotic systems behaving like Axiom A systems there is a link between the presence (or lack)  of regularity  of the parameters describing the extremes of a wide class of observables and of the regularity of the Kaplan-Yorke dimension.  This matches with the fact that numerical practice with moderate to high dimensional strongly chaotic systems shows that, actually, the parameters describing extremes of energy like observables and the Lyapunov exponents both have - within numerical precision - a smooth dependence on the parameters of the system. See \citep{felici1,lucarinietal2007} for an extensive discussion in a simplified yet relevant fluid dynamical model. A detailed investigation of the apparent regularity for all practical purposes of the Lyapunov exponents with respect to small perturbations in the dynamics of intermediate complexity to high-dimensional models has been presented in \cite{albers2006}. We also wish to remark that if these regularity hypotheses were not satisfied, the very widespread (and practically successful) procedure of parametric tuning of high-dimensional models of natural, engineered or social phenomena would be absolutely hopeless, and delicate numerical procedures such as those involved in data assimilation of high-dimensional dynamical systems would lack any sort of robustness, contrary to the accumulated experience.

\section{Numerical Experiments}\label{nume}
As thoroughly discussed in \citep{faranda2011generalized,faranda2011numerical,faranda2011extreme}, it is far from trivial to devise suitable numerical experiments for studying to what extent the theoretically derived asymptotic extreme value laws for dynamical systems can be detected in finite datasets obtained as outputs of simulations. In this Section we would like to present  simple numerical experiments providing some heuristic and preliminary support to the fact that the universal properties for the extremes can be  observed when considering specific observables for dynamical systems. More detailed numerical studies, where inference of the geometrical properties of the attractor is performed using the statistics of extremes of suitable observables will be reported elsewhere. Hence, we consider as \textit{toy model} the smooth $\mathbb{R}^2 \rightarrow \mathbb{R}^2$ map introduced by H\'enon \citep{henon}:
\begin{align}
&x_{n+1}=1-ax_n^2+y_n\\
&y_{n+1}=bx_n.
\label{henon}\end{align}
As well known, depending on the value of the two parameters $a$ and $b$, the H\'enon map can exhibit either regular or chaotic behaviour, where, in the latter case, the invariant measure is supported on a strange attractor  \citep{sprott2003chaos}. We consider two sets of parameter values for which chaotic behavior is observed, $(a,b)=(1.4,0.3)$ and $(a,b)=(1.2,0.3)$. In the first case, the largest Lyapunov exponent $\lambda_1\sim0.416$ and the Kaplan-Yorke dimension is estimated as $d_{KY}=1+\lambda_1/|\lambda_2|=1+\lambda_1/|\log(b)-\lambda_1|\sim1.26$, where $d_u=1$ and $d_s=\lambda_1/|\log(b)-\lambda_1|\sim0.26$. In the second case,  the largest Lyapunov exponent $\lambda_1\sim0.305$ and the Kaplan-Yorke dimension is estimated as $d_{KY}=1+\lambda_1/|\lambda_2|=1+\lambda_1/|\log(b)-\lambda_1|\sim1.20$. Note that it is reasonable to expect that the H\'enon maps considered here do not possess an SRB measure, because the considered pair of values of $a$ and $b$ do not seem to belong to the Benedicks-Carleson set of parameters. As a consequence, these systems are not exact dimensional and the local dimension does not have the same value almost everywhere on the attractor, so, in rigorous terms, we have no a-priori reasons to expect that our results should necessarily apply. Nonetheless, in order to assess the robustness of our findings, it is interesting to check to what extent our theoretical predictions are met, at least qualitatively, in such a basic model of chaotic dynamics. 

We proceed as follows for both pairs of parameters $(a,b)=(1.4,0.3)$ and $(a,b)=(1.2,0.3)$. We first choose, for sake of simplicity, the observable $A(\vec{x})=x$, where $\vec{x}=(x,y)$. The initial conditions are selected in the basin of  attraction of the strange attractors. We perform long integrations (order of $10^{10}$ iterations) and select the maximum value of $A$, which we denote as $A_{max}$, and define as $\vec{x}_0$ the unique point belonging to the attractor such that $A(\vec{x}_0)=A_{max}$. We then construct the observable $B(\vec{x})=-dist(\vec{x},\vec{x}_0)$, which measures the distance between the orbit and the point $\vec{x}_0$. As discussed in \cite{lucarinietal2012}, the asymptotic properties of the extremes (maxima) of the $B$ observable allow to derive easily the local dimension $D(\vec{x}_0)$. We then repeat the investigation using, instead,  the observable $A(\vec{x})=-x$. In all the analyses presented below, we have chosen extremely high thresholds $T$ for studying the statistical properties of the extremes of the $A$ and $B$ observables, in such a way to include only about a fraction of about $10^{-5}$ or less of the total number of points of the orbit. All the results are insensitive to choice of $T$, which suggests that we are well into the asymptotic regime. 

\begin{figure}[t!]
a) \includegraphics[width=0.9\columnwidth]{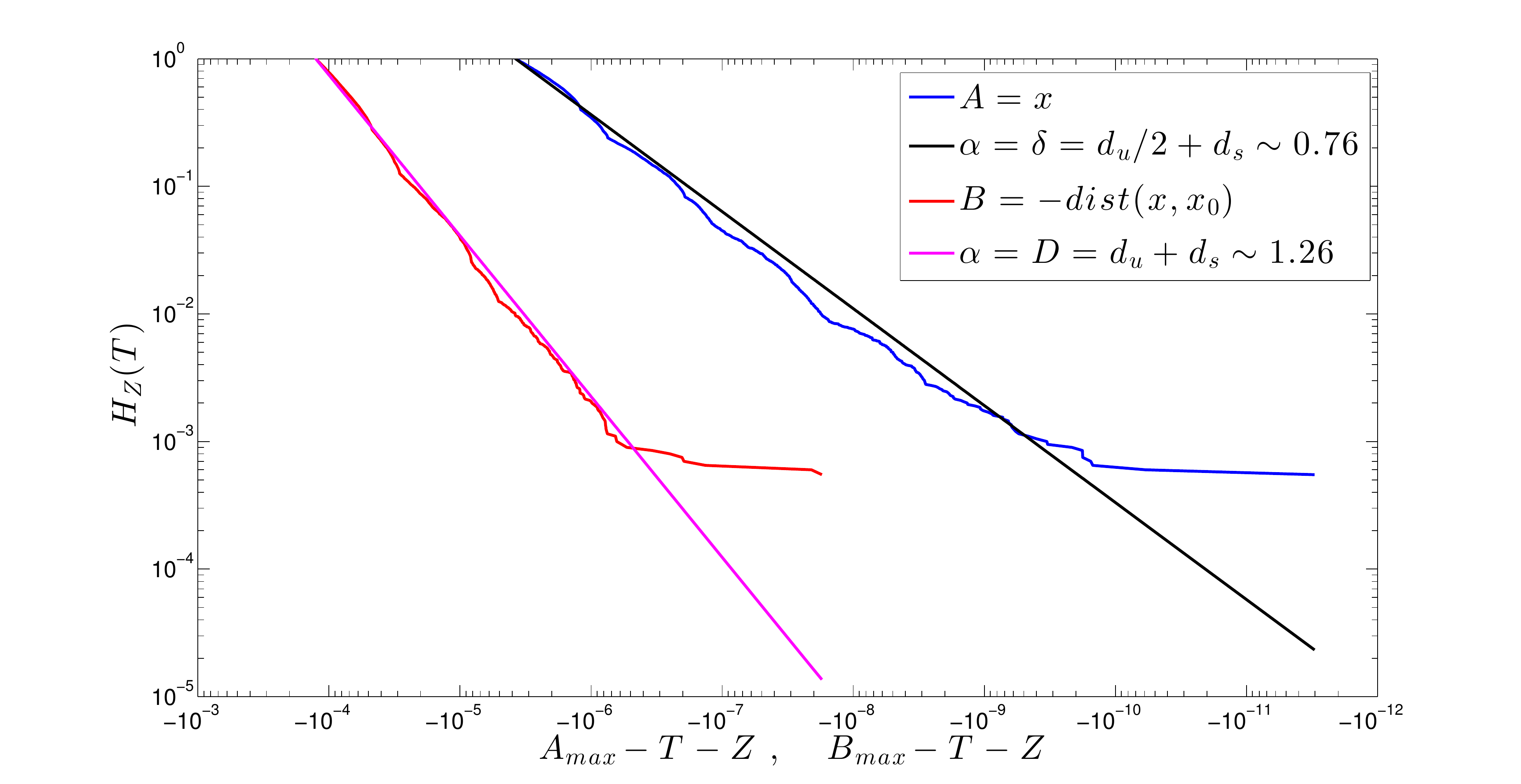}\\
b) \includegraphics[width=0.9\columnwidth]{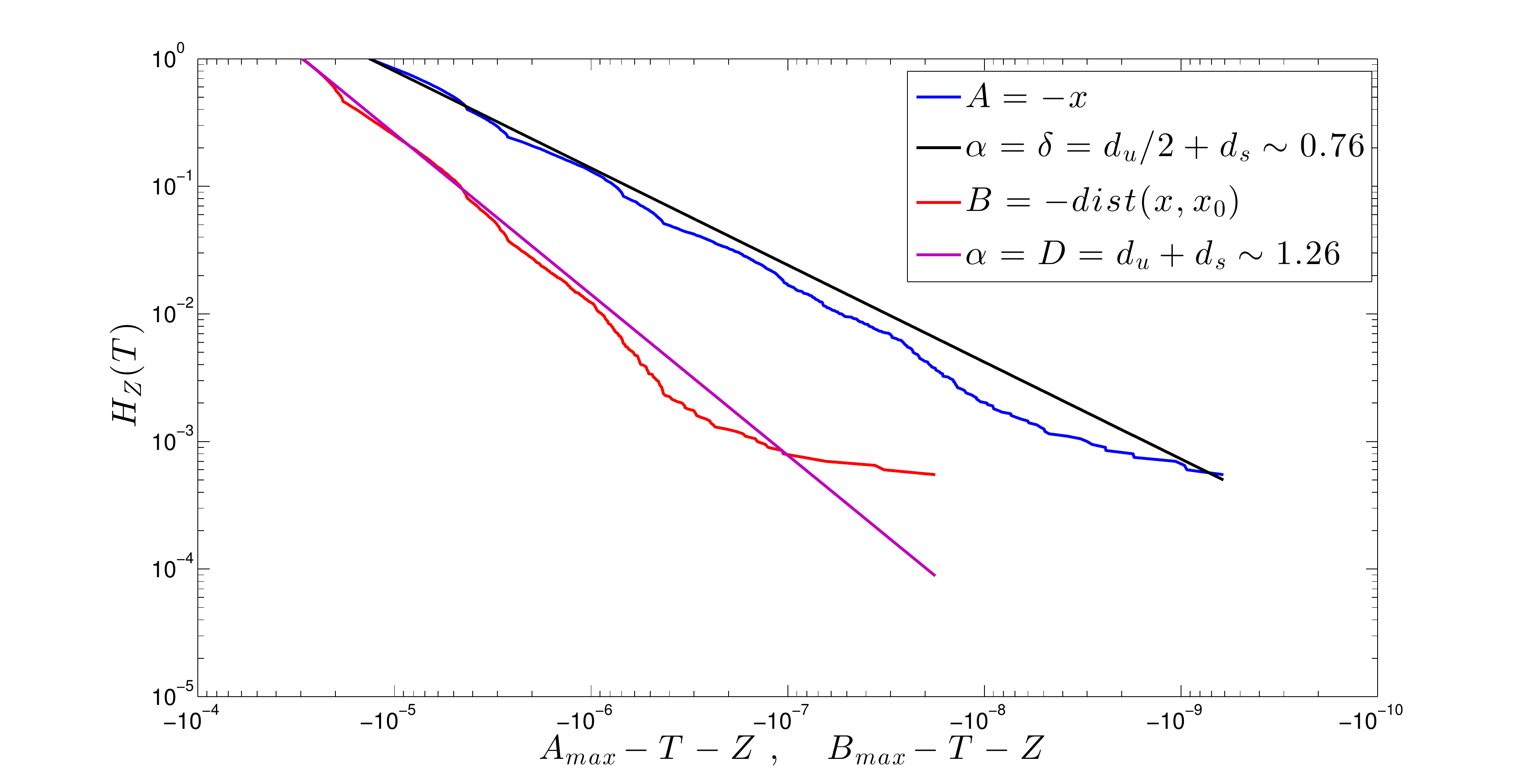}\\
c) \includegraphics[width=0.9\columnwidth]{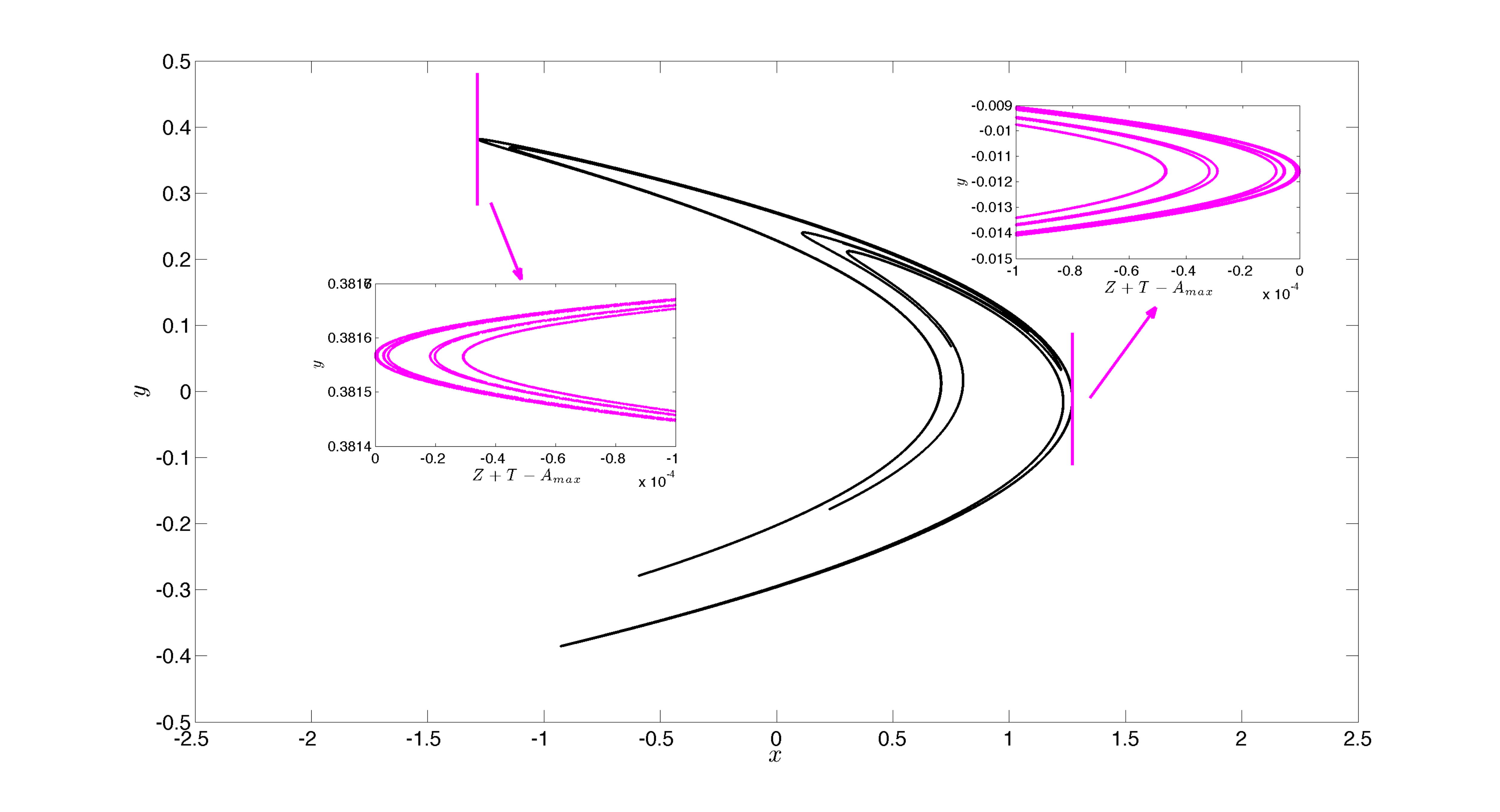}
\caption{Results of numerical simulations performed on the H\'{e}non map with parameters' value $a=1.4$ and $b=0.3$. a) Blue curve: empirical $H_T(Z)$ for the observable $A=x$, with $A_{max}=A(x_0)\sim 1.2730$. Black line: power law behavior deduced from the theory. Red curve: empirical $H_T(Z)$ for the observable $B=-dist(x,x_0)$, with $B_{max}=0$. Magenta line: power law deduced from the theory. b) Same as a), for the observable $A=-x$, with $A_{max}\sim 1.2847$ and $B_{max}=0$. c) Approximation to the attractor with blow-ups of the portions of the invariant measure corresponding to the extremes of the $A$ observables ($\Omega^T_{A_{max}}$ regions); the vertical lines indicate the thresholds. In both inserts, we consider $A_{max}-T=10^{-4}$.  See also Fig. \ref{fig1}.}
\label{fig2}
\end{figure}

The results obtained for the H\'enon  system featuring $(a,b)=(1.4,0.3)$ are shown in Fig. \ref{fig2}, where we present the complementary cumulative distribution of excesses $H_T(Z)$ (see Eqs. \ref{measureball} and \ref{measureball2}) for $A(\vec{x})=x$ ($A(\vec{x})=-x$) and for the corresponding $B(\vec{x})=-dist(\vec{x},\vec{x}_0)$ in panel a) (panel b)). The empirical values of $H_T(Z)$ for the $A$ and $B$ observables are shown by the blue and red curves, respectively, and the power law behavior $H_T(Z)=(1-Z/(A_{max}-T))^\alpha$ given by the theory (assuming Axiom A properties!) are shown by the black and magenta lines, respectively. 

The error bars on the empirical $H_T(Z)$ (estimated by varying the initial conditions of the simulation) are for almost all values of $Z$ so small that they cannot be graphically reproduced. Instead, the flat region obtained for very low values of $H_T(Z)$ results from the finiteness of the sampling and gives the baseline uncertainty. Note that the straight lines are obtained out of  the theoretical predictions, without any procedure of optimization or of fit, so that no uncertainties are involved. The empirical and theoretical distributions obey the same normalization.  

We first observe that the local dimension in the vicinity of both $\vec{x}_0$'s is extremely close to the $d_{KY}\sim 1.26$, as  $H_T(Z)$ scales to a very good approximation with an exponent $\alpha\sim d_{KY}$; compare the red curves and the magenta lines.  Note that, considering that the local dimension has rather large variations across the attractor of the H\'enon system, such a correspondence was not intentionally pursued. However, these are favorable circumstances to check the theory. We find that the distributions $H_T(Z)$ for the observables $A(\vec{x})=x$ and $A(\vec{x})=-x$ also obey accurately the power law scaling with exponent $\alpha\sim\delta=d_u/2+d_s\sim0.76$ introduced in Eq. \ref{deltadef}, compare the blue curves and the black lines.  In panel c) we present a simple description of the geometry of the problem, by showing an approximation to the map's attractor with blow-ups of the portions of the invariant measure corresponding to the extremes of the $A$ observables (the regions $\Omega^T_{max}$ introduced in Fig. \ref{fig1}). Even if the geometrical properties of the regions of the attractor around the two $x_0$'s seem indeed different, when zooming in, the two $\Omega^T_{max}$ regions look  similar. The presence of many parabolas-like smooth curves stacked according to what looks qualitatively like a Cantor set fits with the comments and calculations given in Sec. \ref{EVTAxiomA}.

\begin{figure}[t!]
a) \includegraphics[width=0.9\columnwidth]{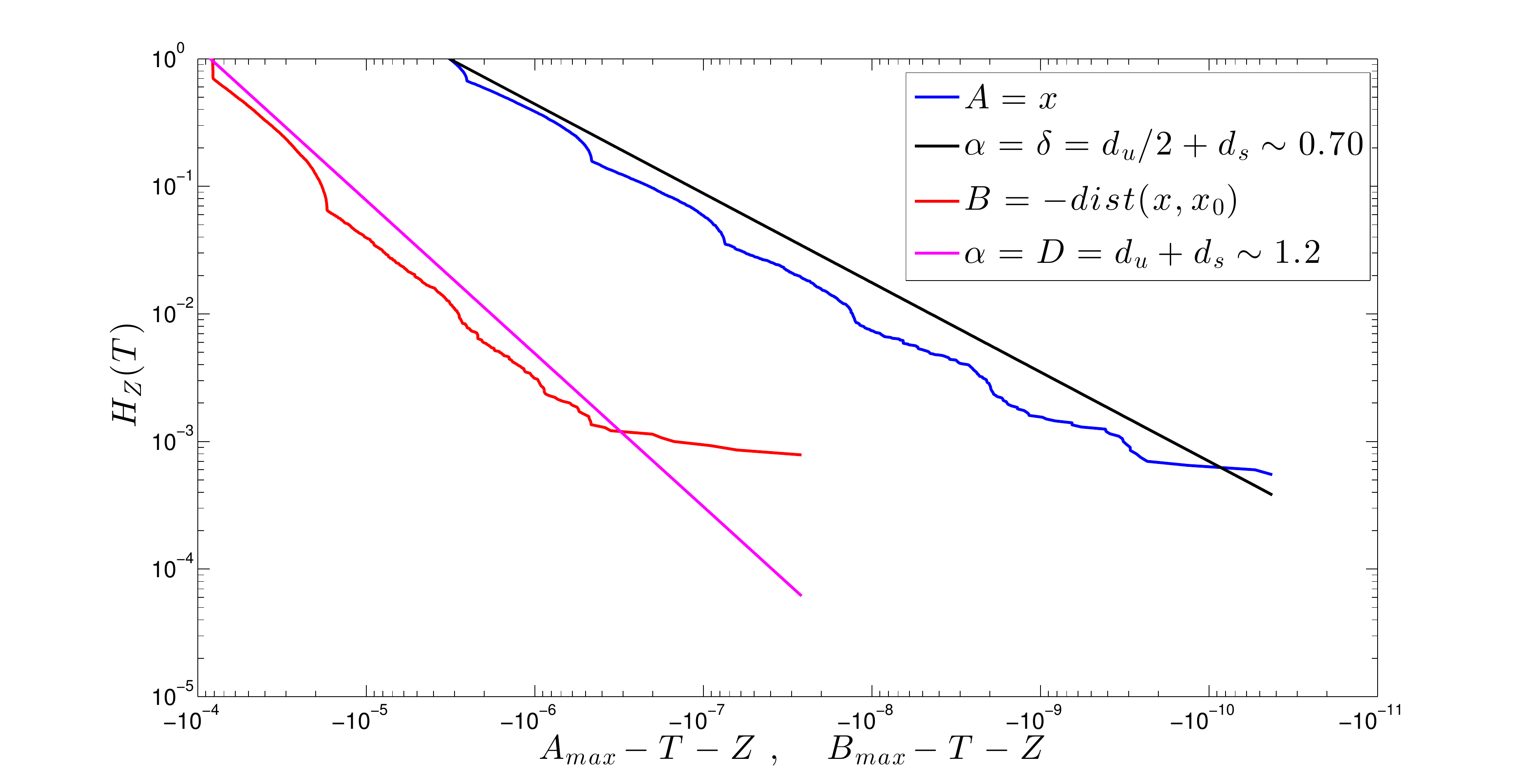}\\
b) \includegraphics[width=0.9\columnwidth]{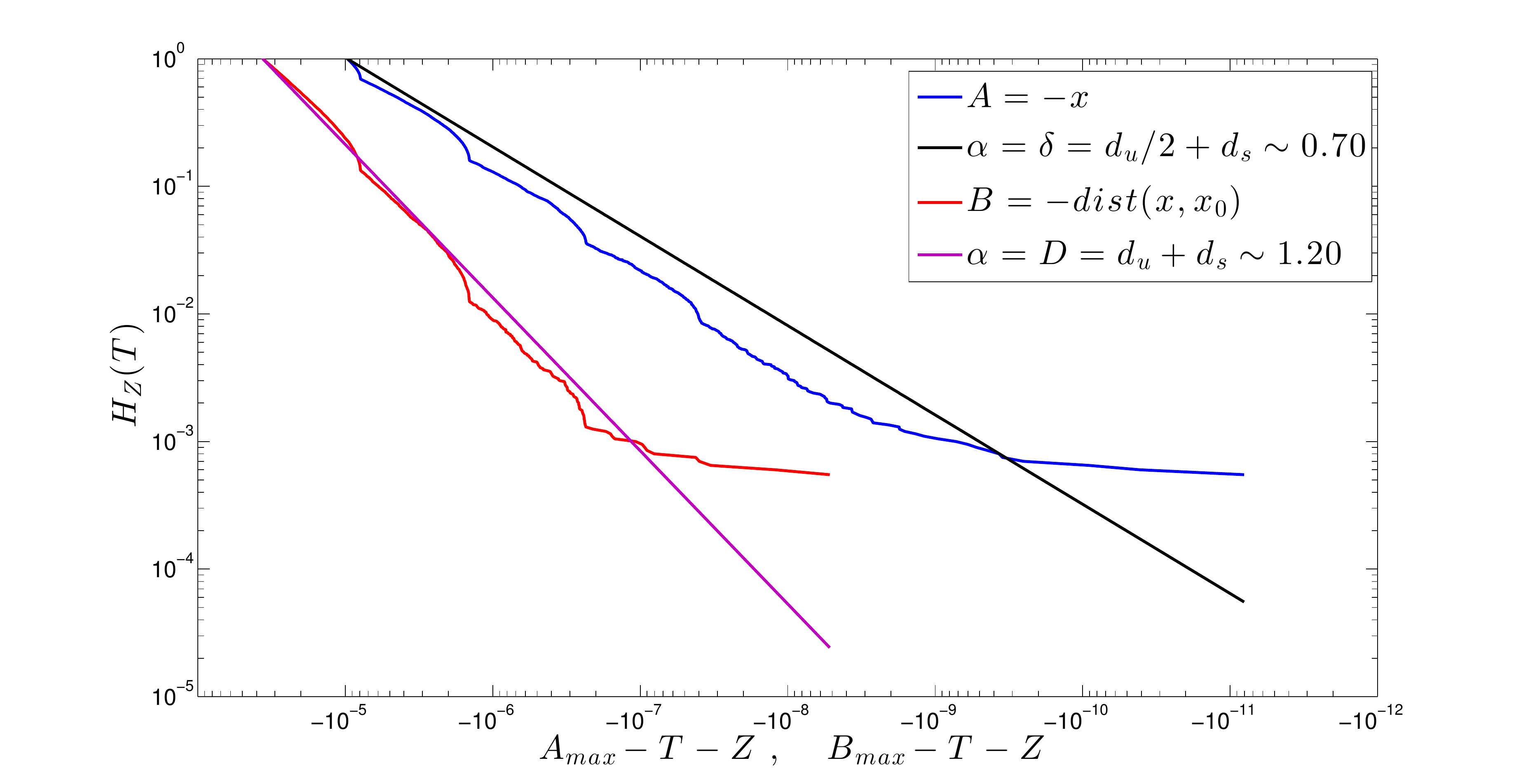}\\
c) \includegraphics[width=0.9\columnwidth]{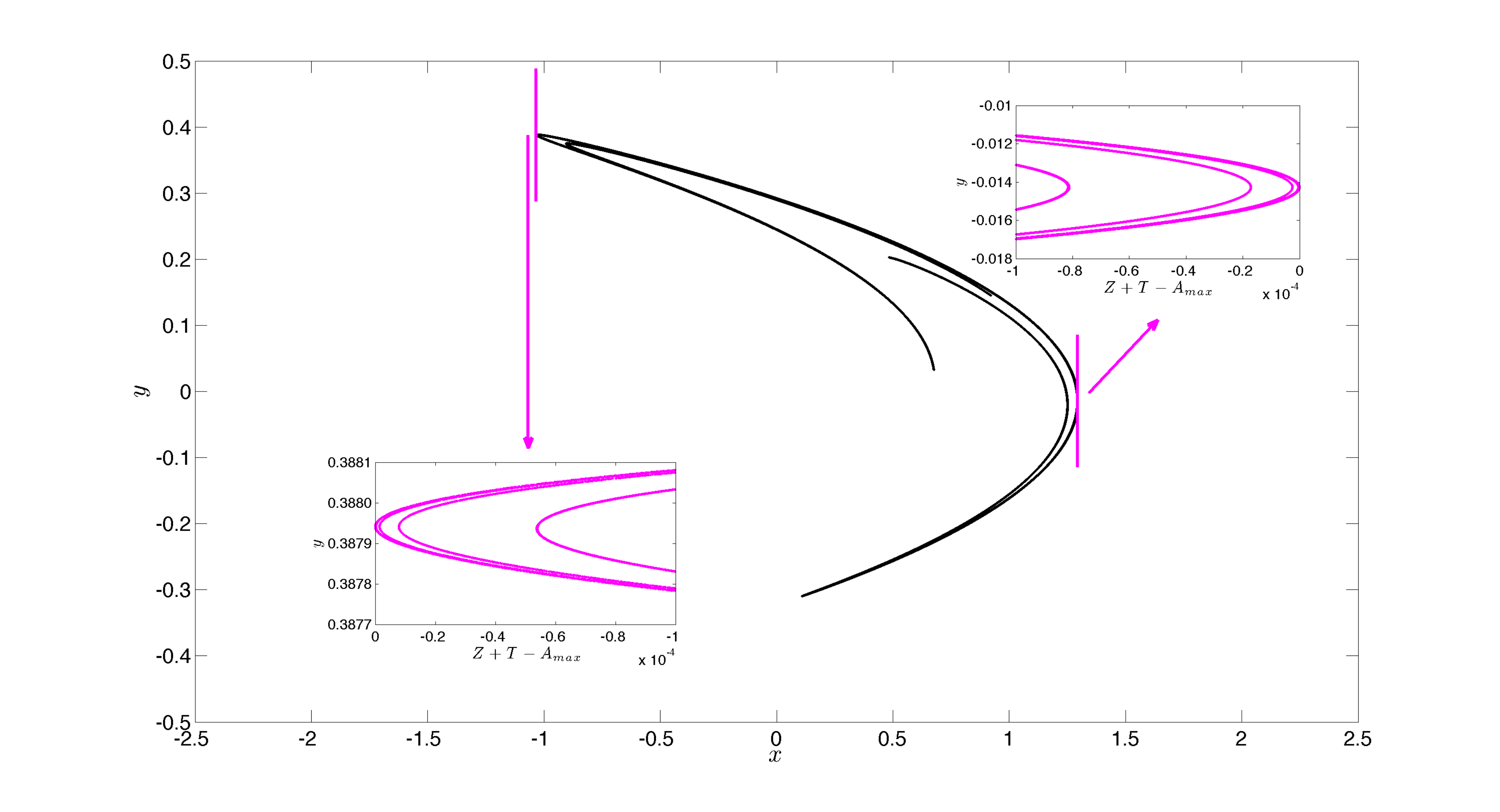}
\caption{Same as in Fig. \ref{fig2}, but for parameters' value $a=1.2$ and $b=0.3$. In this case in a)  $A_{max}\sim 1.2950$ and $B_{max}=0$, and in b) $A_{max}\sim 1.0328$ and $B_{max}=0$.}
\label{fig3}
\end{figure}

In Fig. \ref{fig3} we  report the corresponding results obtained  for the H\'enon  system featuring $(a,b)=(1.2,0.3)$. By looking at the empirical $H_T(Z)$ of the $B$ observables, we note that also in this case the local dimension is close to the value of $d_{KY}\sim1.20$ for both extremal points $x_0$'s (compare the red curves and the magenta lines in panels a and b). Nonetheless, the slope of the empirical data is slightly steeper than the theoretical value. The agreement between the predicted value of the power law scaling for the $H_T(Z)$ of the $A$ observables is not as good as in the case reported in Fig. \ref{fig2}. The predicted scaling exponent $\delta=d_u/2+d_s\sim0.70$ seems to overestimate the very large extremes. Nonetheless, a power law scaling is apparent  for the empirical $H_T(Z)$. Note that the bias between the theoretical and empirical scalings is of the same sign for both the $A$ and $B$ observables, suggesting that also for the $A$ observables part of the disagreement is due to the discrepancy between the local dimension and the Kaplan-Yorke dimension (there is a shift in the values of the slopes). Also here, panel c) provides an approximate representation of the attractor of the system, and, in particular of the $\Omega^T_{max}$ regions: by comparing it with panel c) of Fig. \ref{fig3}, and considering that they contain the same number of points, one can intuitively grasp that the local dimension is lower in this case. 

We would like to emphasize that in panels a) and b) for Figs. \ref{fig2} and \ref{fig3}, we observe deviations of the empirically obtained $H_T(Z)$ from the power law behaviour, in the form of fluctuations above and below a straight line in a $\log$-$\log$ plot (this is quite clear in Fig. \ref{fig3}). As discussed in Sect. \ref{EVTAxiomA} B, the presence of such modulations across scales result from the fact that gaps are present along the stable manifold containing $x_0$, with a Cantor set-like structure. See the inserts in Figs. \ref{fig2}c) and \ref{fig3}c), where the stable manifold (not shown) is, as opposed to the unstable manifold, not orthogonal to the gradient of $A$ (the $x$ direction, in this case). So, when we integrate the density of states along the direction of the gradient of the $A$ observable starting in $\vec{x}_0$ in order to obtain $\nu(\Omega^{T+Z}_{A_{max}})$ and $\nu(\Omega^{T}_{A_{max}})$, we get a  factor $(A_{max}-T-Z)^{d_u/2}$ ($d_u=1$) coming from the (local) paraboloidal form of the unstable manifold discussed in Sec. 2, times  a devil's staircase which can, \textit{on the average}, be approximated by the power law $(A_{max}-T-Z)^{d_s}$. The same geometric arguments apply when considering integrations along the spherical shells centered in $\vec{x}_0$ for constructing the extreme value laws for $B$ observables. Smooth approximation to devil's staircases, appear in a closely related context when using the GEV approach for studying extreme values laws in random dynamical systems whose attractor is the actual Cantor set \citep{faranda2011extreme}. 

Such preliminary results suggest that it is indeed promising to use the combined statistical properties of the extremes of physical and distance observables for determining the geometry of the attractor in terms of its partial dimensions along the stable and the unstable manifold. 


\section{Multiple time scales}
\label{scales}
We briefly wish to mention here some additional features which may appear and be extremely relevant at finite time in  practical cases, where the dynamics can deviate from Axiom A when certain time scales are considered. Let's assume that one can to a first approximation partition the (unique) attractor of a chaotic dynamical system  into, say, two pieces, so that the system has two time scales,  a  short one related to the transitive dynamics within each of the two pieces, and a long one corresponding to intermittent jumps from one to the other piece. In this case, if we observe the system for a time scale intermediate between the two, the properties of the extremes will depend only on the properties of the visited portion of the attractor and we will observe a Weibull distribution, as discussed here, as the dynamics may be Axiom A-like. When our observation time nears the long time scale, we might observe extraordinary large events, corresponding to excursions directed towards the other piece of the attractor, until a jump, corresponding to an irreversible (on the short time scale) transition will take place. Such extraordinary events will not fit the Weibull law found on smaller time scales, because they result from the global properties of the attractor, which have not been sampled yet. Therefore, in these intermediate scale, the results proposed here will not be valid. Instead, one may interpret such extraordinary events as Dragon Kings \cite{sornette2}, which will manifest as outliers spoiling the Weibull statistics and pushing the statistics of extremes towards an (apparently) unphysical Frech\'et distribution. Observing extremes over even longer time scales, so that the orbit visits many times both parts of the attractor, we shall recover a Weibull law, which reflects the global properties of the attractor. In a system with these properties, small perturbations to the dynamics might impact substantially the long time scale discussed above, with the result of having a high sensitivity of the statistics of extremes when a fixed time window of observation is considered. The presence of such strong sensitivity has been proposed as a method for detecting precursors of global stability thresholds related to critical transitions \cite{FLM2013}.

\section{Conclusions}
\label{conclu}

This paper has addressed the problem of studying the EVT for general observables of mixing Axiom A dynamical system. We have set in a common framework the investigation of general properties of \textit{distance observables} $B$, for which we had derived some basic results in \citep{lucarinietal2012}, and  of \textit{physical observables}, first discussed in \citep{holland2012}. By \textit{physical observables} we mean rather general classes of smooth observables $A$ (\textit{e.g.} quadratic, energy-like quantities) which achieve an absolute maximum $A_{max}$ at a specific, non-critical point $x=x_0$ (the gradient $\nabla A$ does not vanish) of the attractor of the system. 

We have built up from the recent results of \citet{holland2012}, who have studied accurately this problem on discrete maps with specific mixing properties using the GEV framework and have derived, as fundamental result, that the extremes of A indeed obey a GEV extreme value law and that $\xi_A$, the shape parameter of the distribution, is equal to $-1/\delta$, with $\delta=d_s+d_u/2$, where $d_s$ is the partial dimension along the stable manifold and $d_u$ is the partial dimension along the unstable manifold. 

In this paper, using the GDP approach, thus considering exceedances above a given threshold $T$, and considering the physically relevant case of mixing Axiom A systems, we derive through direct integration that the shape parameter $\xi_A$ can also be expressed as $\xi_A=-1/\delta$. We have framed our results for continuous flows, so $\delta=d_s+d_u/2+d_n/2$, where $d_n$ is the dimension along the neutral direction and is  unitary. In the case of discrete maps, we obtain the same results as in \cite{holland2012}. We have also been able to derive the explicit expression for the scale parameter $\sigma=(A_{max}-T)/\delta$.

It is clear that the $\xi_A$ parameter is always negative (so that the distribution of extremes is upper limited), reflecting the fact that the observable is smooth and the attractor is a compact set. Moreover, measuring $\xi_A$ allows us to provide an upper and lower bound for $d_{KY}$ and \textit{vice versa}, because $-1/\xi_A<d_{KY}<-2/\xi_A$, or, conversely $d_{KY}/2<-1/\xi_A<d_{KY}$. In particular, we have that $\xi_A$ is small and negative if and only if  the Kaplan-Yorke dimension of the attractor is large. If we consider a chaotic system with a high dimensional attractor (\textit{e.g.} in the case of an extensive chaotic system with many degrees of freedom), we derive that $\xi_A \approx 0$. This may well explain why in a multitude of applications in natural sciences such as hydrology, meteorology, oceanography the special $\xi=0$ member of the GPD family - the exponential model - given in Eq. \ref{GPD} usually gives a good first guess of the statistics of observed extremes  \citep{coles2001introduction}. 

Alternatively, this result suggests that if we perform a statistical analysis using the POT method (using an empirical threshold $T$) of the extremes for a high-dimensional chaotic system and obtain as a result of the statistical inference of the collected data for the GPD model a shape parameter  $\xi_A\ll 0$ or $\xi_A\geq 0$, we should conclude that our sample is not yet suited for an EVT statistical fit. This may depend on the fact that we have selected an insufficiently stringent value for $T$. Obviously, choosing higher values for $T$ implies that we need to have longer time series of the observable under investigations.  

Interestingly, by combining the expression for $\xi_A$ obtained in this paper for a physical observable $A$ and the expression for $\xi_B$ of the GPD describing the extremes of observable of the form $B(x)=C-\text{dist}(x,x_0)^{\beta}$, where dist is the distance function  \cite{lucarinietal2012}, we can express the partial dimension of the attractor on the stable and unstable dimension as simple functions of $\xi_A$ and $\xi_B$. The straightforward result is that $2/\xi_A-2\beta/\xi_B=d_u+d_n$ and $\beta/\xi_B-2/\xi_A=d_s$. The same can be obtained using, instead, the parameters $\sigma_A$ and $\sigma_B$. This provides further support to the idea that extremes can be used as excellent diagnostic indicators for the detailed dynamical properties of a system. The message seems to be that one can construct observables whose large fluctuations give precise information on the dynamics. While  considering various sorts of anisotropic scalings of the neighborhood of a point of the attractor allows to derive its partial dimensions \citep{grassberger1988}, the specific result we obtain here is that choosing an arbitrary \textit{physical} observable and studying its extremes, we automatically select a special, non ellipsoidal neighborhood, where the degree of anisotropy between the stable and unstable (and neutral) directions is generically universal and given by the factor 1/2. 

We wish to make an additional remark. Let's assume, instead, that the gradient of $A$ is vanishing in $x_0$ and that at leading order near $x_0$ $A(x)\sim A_{max}+[x-x_0,H(x-x_0)]$, where $H$ is a negative definite symmetric matrix  and the square brackets indicate the scalar product. It is clear that, apart from a linear change in the coordinates and rescaling, the statistical properties of the extremes of $A(x)$ will match those of $B(x)=C-dist(x,x_0)^2$.

In the second part of the paper we have tackled the problem of studying how the properties of extremes of the observable $A$ change when an $\epsilon$-perturbation is added to the system. As theoretical framework, we have taken the point of view of Ruelle \cite{ruelle98,ruelle2009}, who has shown that the SRB measure of Axiom A systems is differentiable with respect to $\epsilon$-perturbations to the dynamics and has  provided explicit formulas for studying how the expectation values of generic observables of Axiom A systems change when the system is subjected to perturbations. 

We have used the fact that the GPD is an exact asymptotic model for extreme events in order to find a simple functional relation between $\xi_A$ and $\xi_B$ (as well as $\sigma_A$ and $\sigma_B$) and the first two moments of the probability of above-threshold exceedance expressed in terms of the invariant measure of the system. These expressions are amenable to direct treatment with Ruelle's response theory, at least when we do not consider the limit $T\rightarrow A_{max}$ but stick to the practical situation where we need to consider a finite range for the extremes. The differentiability properties of $\xi_A$ and $\xi_B$ are hard to ascertain in the limit. We have also found an explicit expression for the sensitivity of the number of extremes - seen an over threshold events - to the $\epsilon$-perturbation. 

We have then taken into consideration our results on the relationship between $\xi_A$ and $\xi_B$ and the partial dimensions of the attractor.  Interestingly, it seems that there is an intimate connection between the differentiability with respect to $\epsilon$ of $\xi_A$, $\xi_B$ and $d_{KY}$, so that either all of them or none of them is differentiable with respect to $\epsilon$. Under the hypothesis of differentiability, we have been able to derive that the sensitivity of $\xi_A$ and $\xi_B$ with respect to $\epsilon$ is proportional to the sensitivity of the Kaplan-York dimension $d_{KY}$ with respect to $\epsilon$. Specifically, we obtain that if the perturbation tends to increase the dimensionality of the attractor (thus, in physical terms, favoring forcing over dissipation), the value of $\xi$ becomes closer to zero, so that the occurrence of very large extreme events becomes more  likely. The system, in this case, has more freedom to perform large fluctuations. 

Taking a more pragmatic point of view, these results at least provide a rationale for the well-known fact that in moderate to high-dimensional strongly chaotic systems the Kaplan-Yorke dimension (and, actually, all the Lyapunov exponents) change smoothly with the intensity of the perturbating vector field, as discussed in \citep{albers2006,lucarinietal2007}, and similar behavior is found for the parameters describing the extremes of energy-like quantities \citep{felici1}.


Since it has been shown in \citep{lucarini2012} that the linear response vanishes for any observable in the case of stochastic forcing of rather general nature, whereas the second order response gives the leading order of perturbation, we expect that adding a moderate noise to a chaotic dynamical system will not alter significantly the shape parameter $\xi$ describing the EVT of both physical and distance observables.

Finally, we have performed a set of simple numerical experiments using the celebrated H\'enon map for two different pairs of parameters - (a,b)=(1.4,0.3) and (1.2,0.3). While these maps are definitely not Axiom A, it seemed to us worthwhile to test the robustness of the theory in more general classes of systems and to have an indication of whether the asymptotic properties discussed here are practically observable. One has to keep in mind that, when considering extremes, the approach to asymptotic behavior is far from being trivial to detect in finite datasets \citep{faranda2011numerical,faranda2011extreme}. We find encouraging agreement between our theory and the outputs of numerical experiments for both sets of parameters, which suggests that it is worthwhile to study accurately more comprehensive models in order to see whether one can practically derive the geometrical properties of the attractor from the statistics of extremes of distance and physical observables. Further numerical investigations are needed for studying whether it is possible to find satisfactory numerical evidence for the response theory for the extremes developed here and in particular for the relationship between the sensitivity of the EVT's parameters of physical and distance observables and the sensitivity of the Kaplan Yorke dimension to perturbations to the underlying dynamical system. 

In this work we have considered the case where the observable $A$ has a unique maximum restricted to $\Omega$ in $x=x_0\in\Omega$. If $\Omega$ and $A$ share some symmetries, $x_0$ is not unique, and instead there is a set of points $x_0$'s belonging to $\Omega$, finite or infinite, depending of the kind of symmetries involved, where $A$ reaches its maximum value restricted to $\Omega$. Let's consider the relevant case where $A$ and $\Omega$ share a discrete symmetry, so that $\chi_0,$ the set of the maximal point $x_0$'s, has finite cardinality. The results discussed here for the extremes of $A$ will nonetheless apply, because we can perform an equivalent geometrical construction as in Fig. \ref{fig1} for each element of $\chi_0$. When we consider an $\epsilon$-perturbation to the dynamics which respects the discrete symmetry, it is clear that all the results of the response presented here apply. Finally, one can deduce that if the considered perturbation, instead, breaks the discrete symmetry, the results  presented here will still be valid as the break of the degeneracy will make sure that only one of the $x_0$'s (or a subset of $\chi_0$, if the corresponding perturbed vector flow obeys to a a subgroup of the original symmetry group) still accounts for the extreme events of $A$.

This work may constitute a new theoretical viewpoint for studying extremes in a rather general setting and understanding how they change when the dynamical system is slightly perturbed. The results on how to express the sensitivity of the Kaplan-Yorke dimension with respect to an $\epsilon-$perturbation to the dynamics seems also useful. Our findings might find applications in many sectors of physical, engineering and social sciences, and, just to provide a basic example of crucial relevance, in the investigation of the impact of climate change on climate extremes. 

Obviously, as in the case of all results pertaining to EVT, it is important to test numerically the practical verification of the findings presented here. In this paper we have extensively discussed the relevance of finite-time effects in the selection of the extremes of physical observables and in the definition of the relevant sensitivities. We need to mention that, recently, some renormalization group methods have been applied for deriving systematic finite size corrections to extreme value laws \cite{gyorgi1,gyorgi2}. These results seem extremely promising and might lead to improved methods for fitting extreme value statistics to given datasets.

We need to remark that such results have been derived using some intuitive geometrical construction and assuming generic relations between the direction of the gradient of $A$ at $x=x_0$ and the stable directions. It is possible to devise special pair of Axiom A systems and observables such that the strange attractors do not fulfill such generic conditions. One can easily construct a situation where the gradient of $A$ is orthogonal also to stable manifold by immersing the attractor in a higher dimensional space and taking observables defined on such a space. In this case, the factor $1/2$ appearing in Eq. \ref{deltadef} will affect also the stable dimensions. Nonetheless, the results that for high-dimensional systems the distribution of extremes is indistinguishable from the Gumbel as the shape parameters tends to zero from below is not be affected by this correction.  

We believe that typical combinations of Axiom A systems and observable functions allow for the generic conditions to be obeyed. We still need to understand how to frame consistently such a concept of genericity, which obviously differs from the traditional one, which focuses either on the observables, or on the systems. This should be the subject of theoretical investigation and accurate numerical testing.

\begin{acknowledgments}
The authors acknowledge various useful exchanges with J. Freitas, A. M. Freitas, G. Gallavotti, G. G. Gy\"orgyi, M. Holland, Z. R\'acz, T. T\'el, M. Todd, and S. Vaienti, and the comments of two anonymous reviewers. VL, JW and DF acknowledge that the research leading to these results has received funding from the European Research Council under the European Community's Seventh Framework Programme (FP7/2007-2013) / ERC Grant agreement No. 257106, project Thermodynamics of the Climate System - NAMASTE, and has been supported by the Cluster of Excellence CLISAP.  The authors acknowledge the hospitality of the Isaac Newton Institute for Mathematical Sciences (Cambridge, UK) during the 2013 programme \textit{Mathematics for the Fluid Earth}. 
\end{acknowledgments}
\end{document}